# Mapping Equivalence for Symbolic Sequences: Theory and Applications

Liming Wang, *Student Member, IEEE,* and Dan Schonfeld, *Senior Member, IEEE*

**Abstract**

Processing of symbolic sequences represented by mapping of symbolic data into numerical signals is commonly used in various applications. It is a particularly popular approach in genomic and proteomic sequence analysis. Numerous mappings of symbolic sequences have been proposed for various applications. It is unclear however whether the processing of symbolic data provides an artifact of the numerical mapping or is an inherent property of the symbolic data. This issue has been long ignored in the engineering and scientific literature. It is possible that many of the results obtained in symbolic signal processing could be a byproduct of the mapping and might not shed any light on the underlying properties embedded in the data. Moreover, in many applications, conflicting conclusions may arise due to the choice of the mapping used for numerical representation of symbolic data. In this paper, we present a novel framework for the analysis of the equivalence of the mappings used for numerical representation of symbolic data. We present strong and weak equivalence properties and rely on signal correlation to characterize equivalent mappings. We derive theoretical results which establish conditions for consistency among numerical mappings of symbolic data. Furthermore, we introduce an abstract mapping model for symbolic sequences and extend the notion of equivalence to an algebraic framework. Finally, we illustrate our theoretical results by application to DNA sequence analysis.

The authors are with the Department of Electrical and Computer Engineering, University of Illinois at Chicago, Chicago, IL 60607-7053 USA (e-mail: lwang37@uic.edu; dans@uic.edu)

# I. Introduction

Information is provided in many forms. At times, information is conveyed numerically. More often, information is represented in the form of symbols such as characters, tags, etc. For example, the areas of genomic and proteomic signal processing focus on sequences of nucleotides and amino acids, respectively [1]. The aim of symbolic signal analysis is to process symbolic data elements in order to extract useful information.

In general, symbolic information is represented as a sequence of symbols (possibly of infinite length) $\{a_i\}_{i=0}^{N-1}$, where $a_i \in \mathcal{A}$ and $\mathcal{A}$ is a set of all possible symbols. For example, $\mathcal{A}$ could be a collection of the 26-lowercase English letters, i.e. $\mathcal{A} = \{a,\ b, \ldots,\ z\}$, or the four nucleotides in a genomic sequence, i.e. $\mathcal{A} = \{\mathtt{A},\ \mathtt{T},\ \mathtt{G},\ \mathtt{C}\}$. In statistical literature, symbolic data is usually called categorical data [2]. The use of Markov chain models and hidden Markov models has been examined for time-domain analysis of genomic and proteomic data [3], [4], [5], [6]. However, we often seek to rely on frequency-domain analysis methods of symbolic signals. Unfortunately, symbolic sets do not generally possess an algebraic structure that allows us to define mathematical operations (e.g. group, ring, or field). In traditional signal processing, the set $\mathcal{A}$ corresponds to real- or complex-valued numbers, i.e. $\mathcal{A} = \mathbb{R}\ or\ \mathbb{C}$, which form an algebraic field. However, attempts to define mathematical operations such as addition and multiplication on symbolic data has raised many questions about the meaning of the results obtained using such methods.

Several techniques exist which incorporate numerical and symbolic processing in an effective way to develop symbolic analysis systems [7]. Software systems for symbolic computational algebra (e.g. Mathematica, Maple, etc.) represent a successful example of this approach. Such systems, however, are application-specific and difficult to realize for a broad class of symbolic signal processing applications. There are also various techniques for analyzing correlations, periodicities, etc. that do not require the aid of numerical symbol mappings. Among these techniques, the Mutual Information Function (MIF) [8] is one of the most important. The main advantage of these methods is that numerical mappings are not required. Moreover, it can be shown that methods such as MIF can capture any type of statistical dependence. The main disadvantages of these techniques, however, are that they generally provide less specific information than correlation analysis and they often suffer from a systematic overestimation of mutual information for finite sequences. Nevertheless, in order to extract the mathematical and statistical information embedded in symbolic sequences, we wish to employ the powerful analysis tools developed in traditional signal processing, e.g. Fourier transform, correlation function, etc. We must therefore map the symbolic elements into numerical values. The resulting numerical sequence should preserve the information embedded in the symbolic data. Moreover, it should allow traditional signal processing techniques to extract the salient information about the symbolic sequences from the corresponding numerical signals. For instance, in DNA sequences, we have a finite alphabet associated with the four nucleotides in the genome, i.e. $\mathcal{A} = \{\texttt{A}, \texttt{T}, \texttt{G}, \texttt{C}\}$. The mapping used for the representation of genomic data must preserve the inherent structure of DNA sequences. In particular, if we choose a mapping such as: $\texttt{A} \mapsto 1$, $\texttt{T} \mapsto 0$, $\texttt{G} \mapsto -1$, $\texttt{C} \mapsto 0$, we would not preserve uniqueness since T and C are mapped to the same value.

Numerous mappings have been proposed for the numerical representation of DNA sequences. Buldyrev et al. [9] proposed various mapping rules for the representation of nucleotide sequences into one-dimensional numerical sequences based on the purine-pyrimidine (RY) rule, hydrogen-bond energy rule, etc. Li and Kaneko [10] and Voss [11] used the indicator sequence method, which essentially maps the symbol to a standard basis of the 4-dimensional Euclidean space $\mathbb{R}^4$. Berthelsen et al. [12] revised the method introduced in [9] by taking the molecular mass and hydrophobicity into account in representation of genomic data. Silverman and Linsker [13] relied on the simplex method, which maps the symbol to the vertices of a regular simplex. Cristea and Anastassiou [14], [15] proposed the tetrahedral mapping, which maps the nucleotides into corners of a tetrahedron. Stoffer et al. [16] introduced a mapping whose aim is to accentuate the periodic features embedded in genomic sequences for stationary symbolic sequence analysis. Wang and Johnson [17] extended the method proposed by Stoffer et al. [16] for non-stationary sequence analysis. Rushid and Tuqan [18] proposed the Z-curve mapping, which is a unique 3-dimensional curve representation whose sequences are composed of binary values, i.e. $1$ and $-1$. They also proposed a matrix-based framework to combine many widely used mapping strategies in genomic sequence analysis [19]. Akhtar and Epps [20] proposed the Paired Numeric and Frequency of Nucleotide Occurrence methods for DNA symbolic-to-numeric representation and greatly improved the relative accuracy for gene and exon prediction. Asif and Datta [21] developed theoretical properties for the Binary Indicator Sequence method. Tuqan and Rushid [22] proposed a new DSP approach for finding codon bias based on Voss Indicator Sequence method.

Each of the large number of numerical mappings used for the representation of genomic sequences can be justified for various applications. This raises several fundamental questions: What are the merits of each mapping used for the analysis of DNA sequences? How can we compare the results obtained from different numerical mappings? Indeed, it is impossible to determine which mapping is preferable. Furthermore, it is conceivable that distinct mappings could lead to contradictory conclusions. In fact, several contradictory results have arisen in the field of genomic sequence analysis. Most notably, the study of long-range correlations in coding and non-coding DNA sequences has been contested by several contradictory results [11], [23], [24], [25]. Investigation using a large DNA sequence database did not resolve this dispute; in fact, the controversy grew even further [26]. Bouaynaya and Schonfeld [27], [28] shed light on this dilemma by demonstrating that a certain class of genomic sequences are inherently non-stationary and thus one of the reasons for the contradictory conclusions stems from the use of stationary time-series analysis tools. Moreover, they determined experimentally that the results obtained remained invariant over a large class of numerical mappings used for the representation of DNA sequences. Nonetheless, the experimental study conducted by Bouaynaya and Schonfeld in [27], [28] cannot be used to ascertain with certainty whether the different numerical mappings used for representation of genomic sequences contributed to the contradictory findings reported in the literature [11], [23], [24], [25].

To ensure a clear understanding of the implications of the different choices used for numerical representation of symbolic data, we must develop a fundamental new approach that can be used to characterize the fundamental properties of numerical mappings. Specifically, it is essential that we establish a mapping equivalence theory for symbolic data that can be used to guarantee consistency among a class of numerical representations. With the aid of a mapping equivalence theory we could determine whether different mappings should yield compatible results, i.e. whether the mappings used for the analysis of the same data lead to consistent conclusions. Moreover, the theory can indicate when distinct mappings could lead to contradictory results and thus comparison of the corresponding conclusions is futile.

In this paper, we provide a mapping equivalence theory for the numerical representation of symbolic data undergoing transformation by an operator. We focus primarily on the mapping $f : \mathcal{A} \to \mathbb{R}^n$ which maps the symbols to the $n$-dimensional Euclidean space. In Section II, we first propose a framework for the analysis of different numerical mappings undergoing transformation by an analytic operator using Taylor's expansion. Moreover, we emphasize the investigation of first- and second-order operators including the correlation function and Fourier transform. These operators are widely used in signal processing and analysis and thus play an important role in this presentation. In Section II-A, we provide an analysis of the correlation between different numerical mappings of a symbolic sequence. In particular, we derive conditions for strong equivalence captured by perfect correlation among distinct mappings. In Section II-B, we explore a relaxed similarity measure between distinct numerical mappings. Specifically, we provide conditions for weak equivalence which is characterized by preservation of the local extrema of the representation. In Section III, we introduce an abstract mapping model and extend the concept of equivalence to the generalized mapping model. In Section IV, we present experimental results which illustrate the significance of the proposed mapping equivalence theory in symbolic signal processing applications. In this presentation, the simulations are focused exclusively on analysis of genomic sequences. The results presented in this paper, however, are applicable for any symbolic signal modeled by a discrete alphabet with a finite cardinality and independent of particular statistical properties such as stationarity, etc. Finally, we provide a brief summary and discussion of our results in Section V.

## II. Euclidean Mapping Equivalence for Symbolic Sequences

Given $\{a_i\}_{i=0}^{N-1}$, where $a_i \in \mathcal{A}$ and $|\mathcal{A}| < \infty$, here $|\cdot|$ means the cardinality of the set. $f$ is a mapping from $\mathcal{A}$ to $\mathbb{R}^n$, i.e. $f : a_i \mapsto x_i,\ x_i \in \mathbb{R}^n$. After the mapping we obtain a vector sequence $\{x_i\}_{i=0}^{N-1}$. $T : x_i \mapsto y_i$ is a transformation from $\mathbb{R}^n$ to $\mathbb{R}^m$. $\mathbf{\Phi}_l$ is an analytic operator on the numerical sequence and maps into $\mathbb{R}$ parameterized by $l \in \mathbb{R}$. We also assume that $\mathbf{\Phi}_l \in L^2(l)$. We classify the problems as in the following cases.

1) Given $T$, determine the consistency between $\mathbf{\Phi}_l(\{x_i\}_{i=0}^{N-1})$ and $\mathbf{\Phi}_l(\{T(x_i)\}_{i=0}^{N-1})$. We need also figure out the largest class of operators which shows consistent results for two mappings under the given $T$.
2) Given $f$ and $\mathbf{\Phi}_l$, if $f$ and $T \circ f$ are consistent for any symbolic sequence $\{a_i\}_{i=0}^{N-1}$. Find out the largest class of such transformation $T$ which preserves the consistency. Also figure out the largest class of transformation $T$ preserving the consistency for given mapping $f$.

The consistency here means we require the results under two different mappings to be similar in certain extent. In general $\mathbf{\Phi}_l$ may not be linear. We will use Taylor's expansion to expand the operator. We vectorize the vector sequence $\{x_i\}_{i=0}^{N-1}$ $x_i \in \mathbb{R}^n$ to a large vector $x \in \mathbb{R}^{Nn\times 1}$. Consider the Taylor's expansion of the analytic operator. $\mathbf{\Phi}_l : \mathbb{R}^{Nn\times 1} \to \mathbb{R}$. Without using the common scalar form representation of Taylor's expansion [29], we shall present it in a concise form by using tensor product. First, we define the gradient operator $\nabla$ as

$$\nabla = \begin{pmatrix} \frac{\partial}{\partial x_1} & \frac{\partial}{\partial x_2} & \cdots & \frac{\partial}{\partial x_{Nn}} \end{pmatrix}^T \tag{1}$$

Then the Taylor's expansion of $\mathbf{\Phi}_l$ at $x_0$ can be represented as the following form,

$$\mathbf{\Phi}_l = \sum_{i=0}^{\infty} \frac{1}{i!} (\nabla^i \mathbf{\Phi}_l)(x_0) \times_i (X - x_0) \times_{i-1} (X - x_0) \times_{i-2} \cdots \times_1 (X - x_0) \tag{2}$$

Where $\times_i$ is the $i^{th}$ mode tensor product [30], and $\nabla^i$ is the $i^{th}$ order gradient of $\mathbf{\Phi}_l$, which is defined as,

$$\nabla^i \mathbf{\Phi}_l = \mathbf{\Phi}_l \times_1 \nabla \times_2 \nabla \times_3 \cdots \times_i \nabla \tag{3}$$

Furthermore, $\nabla^0 \mathbf{\Phi}_l$ is defined as $\mathbf{\Phi}_l$. For one- and second-order terms, it is easy to check that it coincides with the well-known definition of *Gradient* $\nabla \mathbf{\Phi}_l(x)$ and *Hessian* $\nabla^2 \mathbf{\Phi}_l(x)$. So we can rewrite the Taylor's expansion at $x_0$ as,

$$\begin{aligned} \mathbf{\Phi}_l =& \mathbf{\Phi}_l(x_0) + \nabla \mathbf{\Phi}_l(x_0)^T (x - x_0) + \frac{1}{2}(x - x_0)^T \nabla^2 \mathbf{\Phi}_l(x_0)(x - x_0) \\ &+ \sum_{i=3}^{\infty} \frac{1}{i!} (\nabla^i \mathbf{\Phi}_l)(x_0) \times_i (x - x_0) \times_{i-1} (x - x_0) \times_{i-2} \cdots \times_1 (x - x_0) \end{aligned} \tag{4}$$

A metric or measure is needed for measuring the consistency. In general, there is no universal metric. Various operators may have different metrics for different purposes. In many cases, it is a reasonable principle to require the results of two different mappings to be similar in some extent. In light of this principle, we propose the following two kinds of metrics.

### *A. Strong Equivalence: Perfect Correlation*

We will use the correlation coefficient to characterize the consistency. First we provide the definition of the correlation coefficient $\rho$ used in this paper.

*Definition 1:* Given $\{a_i\}_{i=0}^{N-1}$, where $a_i \in \mathcal{A}$, $|\mathcal{A}| < \infty$. $f : a_i \mapsto x_i,\ x \in \mathbb{R}^n$, $T : x_i \mapsto y_i$ is a transformation from $\mathbb{R}^n$ to $\mathbb{R}^m$, $\mathbf{\Phi}_l$ is an operator on the numerical sequence. $m(\mathbf{\Phi}_l) = \frac{1}{\mu(L)} \int_l \mathbf{\Phi}_l d\mu$ is the mean value of the $\mathbf{\Phi}_l$ in the space $L$ of parameter $l \in \mathbb{R}$. $\mu$ is a measure on $\mathbb{R}$. The *correlation* *coefficient* is defined as

$$\rho = \frac{\int_l [\mathbf{\Phi}_l(\{x_i\}_{i=0}^{N-1}) - m(\mathbf{\Phi}_l(\{x_i\}_{i=0}^{N-1}))]}{\sqrt{\int_l (\mathbf{\Phi}_l(\{x_i\}_{i=0}^{N-1}) - m(\mathbf{\Phi}_l(\{x_i\}_{i=0}^{N-1})))^2 d\mu}} \frac{[\mathbf{\Phi}_l(\{T(x_i)\}_{i=0}^{N-1}) - m(\mathbf{\Phi}_l(\{T(x_i)\}_{i=0}^{N-1}))] d\mu}{\sqrt{\int_l (\mathbf{\Phi}_l(\{T(x_i)\}_{i=0}^{N-1}) - m(\mathbf{\Phi}_l(\{T(x_i)\}_{i=0}^{N-1})))^2 d\mu}} \tag{5}$$

The use of abstract integration provides a unified framework for definition of the correlation coefficient. The measure $\mu$ can be chosen to be any Borel measure such as the Lebesgue-Stieltjes or counting measures depending on the properties of the operator. In practice, the measures we rely upon are mainly the counting measure and the Lebesgue measure.

It is well known that the correlation coefficient is between $[-1, 1]$ [31]. The correlation coefficient can be used as a measure to characterize the similarity of two different mappings. For a given $T$, if $\rho = 1$, then we say the transformation $T$ is a *strongly equivalent* transformation of the map $f$ for an operator and $\mathbf{\Phi}_l(\{T(x_i)\}_{i=0}^{N-1})$ is a *strong equivalence* of $\mathbf{\Phi}_l(\{x_i\}_{i=0}^{N-1})$. When the correlation coefficient is 1, it means the two mappings are the same up to a translation and scaling. This is the reason that it is called "strongly equivalent." Unfortunately, there is no the universal equivalent transformation for arbitrary operator. However, because of the importance of second-order statistics, we shall emphasize on the second-order operators such as the correlation function. From now on we will focus on the transformation $T$ from $\mathbb{R}^n$ to $\mathbb{R}^n$. For the case of mapping between Euclidean spaces with a different dimensions, we will present a detailed discussion in Section II-D.

We first consider the correlation function. The correlation function of a sequence is defined as,

$$r_l = \frac{1}{N} \sum_{n=0}^{N-1} x^T(n) x(n-l) \tag{6}$$

Then if $\rho = 1$, we have the following theorem on the strongly equivalent transformation $T$.

*Theorem 1:* For non-trivial operator and linear transformation $T$, the correlation coefficient $\rho = 1$ if and only if the transformation $T$ can be represented as $T(x_i) = \lambda \mathbf{R} x_i$, $\mathbf{R}$ is an orthogonal matrix and $\lambda \in \mathbb{R}$.

*Proof:* If $T(x_i) = \lambda \mathbf{R} x_i$, and $\mathbf{R}$ is orthogonal. Then

$$\begin{aligned} r_l(T(\{x_i\}_{i=0}^{N-1})) &= \frac{1}{N}\sum_{n=0}^{N-1} \lambda x_i^T \mathbf{R}^T \mathbf{R} x(n-l) \\ &= \frac{1}{N}\sum_{n=0}^{N-1} \lambda x_i^T x(n-l) \\ &= \lambda r_l(\{x_i\}_{i=0}^{N-1}) \end{aligned} \tag{7}$$

Conversely, if $\rho = 1.$ and $T(x_i) = \mathbf{A} x_i$. Then

$$\begin{aligned} r_l(T(\{x_i\}_{i=0}^{N-1})) &= \frac{1}{N}\sum_{n=0}^{N-1} x_i \mathbf{A}^T \mathbf{A} x(n-l) \\ &= \lambda' r_l(\{x_i\}_{i=0}^{N-1}) + c \\ &= \lambda^{'} \frac{1}{N}\sum_{n=0}^{N-1} x^T(n) x(n-l) + c \end{aligned} \tag{8}$$

where $\lambda^{'} \in \mathbb{R}$ and $c \in \mathbb{R}$ is a constant. Since the equality holds for any sequence and any $l$. So $\mathbf{A}^T \mathbf{A} = \lambda I$, then $\mathbf{A}$ is orthogonal. ■

Actually, this property not only holds for correlation operator, but also for a larger class of operators. Consider the Taylor's expansion of an operator $\mathbf{\Phi}_l$. We would like first to introduce the definition of bounded linear operator and Riesz representation theorem [32]. Then we will present a result for the first- and second-order bounded operators.

*Definition 2:* Let $(X, ||\cdot||)$ be a normed space. A operator $f : X \to \mathbb{R}$ is a bounded operator if $f$ is linear and there exists $C > 0$, such that $|f| \leq C\|x\|$.

The bounded operator can be thought as an analog of BIBO linear system in signal processing theory, which illustrates the good behaved operators. Furthermore if the space $X$ is a Hilbert space, we have the following theorem to characterize any linear bounded operator.

*Theorem 2:* (Riesz Representation theorem for Hilbert space) $X$ is a Hilbert space, then for any linear bounded operator $\phi$, there exists a unique $y \in X$, such that $\phi(x) =< x, y >$.

Note that $\mathbb{R}^n$ with the usual dot product is a Hilbert space. Therefore, Riesz Representation theorem for Hilbert space holds for $\mathbb{R}^n$. As before, we vectorize the vector sequence $\{x_i\}_{i=0}^{N-1}$ $x_i \in \mathbb{R}^n$ to a large vector $x \in \mathbb{R}^{Nn\times 1}$. Then any linear transformation $T$ can be represented in the form

$$T = \begin{pmatrix} \mathbf{A}_{n\times n} & & & \\ & \mathbf{A}_{n\times n} & & \\ & & \ddots & \\ & & & \mathbf{A}_{n\times n} \end{pmatrix}_{Nn\times Nn} \tag{9}$$

i.e. $y = Tx$.

Then we have the following theorem for equivalent transformation of first- and second-order operators.

*Theorem 3:* Any non-trivial bounded linear operator does not have any non-trivial (scaled identity mapping) linear *strongly equivalent* transformation. If rotation is a *strongly equivalent* transformation for a bounded operator whose Taylor's expansion does not have the third or higher-order terms, then its Taylor's expansion can not have first-order term and the Hessian $\nabla^2\mathbf{\Phi}_l(x)$ must have the form

$$\nabla^2\mathbf{\Phi}_l(x) = \begin{pmatrix} k_{11}I_{n\times n} & k_{12}I_{n\times n} & \cdots & k_{1N}I_{n\times n} \\ k_{21}I_{n\times n} & k_{22}I_{n\times n} & \cdots & k_{2N}I_{n\times n} \\ \vdots & \vdots & \ddots & \vdots \\ k_{N1}I_{n\times n} & k_{N2}I_{n\times n} & \cdots & k_{NN}I_{n\times n} \end{pmatrix} \tag{10}$$

, where $k_{ij} \in \mathbb{R}$ and $k_{ij} = k_{ji}$, $\forall i \neq j$.

For Fourier transform, in many situations, we focus exclusively on the modulus of the transform of symbolic data, i.e. we discard the phase information. Since the module of continuous-time Fourier transform is invariant under rotation, it is tempting to conclude that rotation is an equivalent transformation for the Fourier transform. However, the widely used form of the Fourier transform used in much of the literature devoted to DNA sequence analysis [17] is different from the classical multi-dimensional Fourier transform. Fortunately, we are able to show that rotation still yields an equivalent transformation. We first define the Fourier transform as:

$$\hat{f}_m = \frac{1}{N^2}||X\mathbf{L_F}||_2^2 \tag{11}$$

where $X$ is a $n \times N$ matrix, whose $i^{th}$ column is $x_i$. $\mathbf{L_F}$ is the frequency vector, i.e.

$$\mathbf{L_F} = \begin{pmatrix} e^{\frac{-2\pi jm0}{N}} & e^{\frac{-2\pi jm1}{N}} & e^{\frac{-2\pi jm2}{N}} & \cdots & e^{\frac{-2\pi jm(N-1)}{N}} \end{pmatrix}^T \tag{12}$$

If we vectorize $X$ to $x \in \mathbb{R}^{Nn\times 1}$ as before, $\hat{f}_m$ can also be represented as $\hat{f}_m = \frac{1}{N^2}||Ax||_2^2$, where

$$A = \begin{pmatrix} e^{\frac{-2\pi jm0}{N}} I_{n\times n} & e^{\frac{-2\pi jm1}{N}} I_{n\times n} & e^{\frac{-2\pi jm2}{N}} I_{n\times n} & \cdots & e^{\frac{-2\pi jm(N-1)}{N}} I_{n\times n} \end{pmatrix}_{n\times N} \tag{13}$$

notice that $\hat{f}_m = \frac{1}{N^2}(x^H A^H A x)$, which is a second-order operator and $A^H A$ is of the form,

$$A^H A = \begin{pmatrix} \bar{w}_0 w_0 I_{n\times n} & \bar{w}_0 w_1 I_{n\times n} & \cdots & \bar{w}_0 w_N I_{n\times n} \\ \bar{w}_1 w_0 I_{n\times n} & \bar{w}_1 w_1 I_{n\times n} & \cdots & \bar{w}_1 w_N I_{n\times n} \\ \vdots & \vdots & \ddots & \vdots \\ \bar{w}_N w_0 I_{n\times n} & \bar{w}_N w_1 I_{n\times n} & \cdots & \bar{w}_N w_N I_{n\times n} \end{pmatrix} \tag{14}$$

here $w_i = e^{\frac{-2\pi jmi}{N}}$, By theorem 3, rotation is a strongly equivalent transformation for Fourier transform.

### *B. Weak Equivalence: Preservation of Local Extrema*

In the previous section, we employed the correlation coefficient as a metric to characterize the similarity for an operator under transformation. However, as we can see, the *strong equivalence* basically requires the result to be “exactly” the same. While in many situations, we do not focus on whether or not the result under two mapping strategies are exactly the same, i.e. the true numerical value of the result, but the relative relation or the relative trend of the result. For example, when we use the correlation function, in many cases, we only care where the peak and valley points are located and the changing trends, which are used to determine the periodicity structure of certain patterns. In these cases, what we really need is to preserve the local extremums and local trend under the a transformation. So we first give the definition of Local Minimum and Maximum Preserving Similarity or in this paper what we call *weakly equivalent*.

*Definition 3:* Given $\{a_i\}_{i=0}^{N-1}$, where $a_i \in \mathcal{A}$, $|\mathcal{A}| < \infty$. $f : a_i \mapsto x_i,\ x \in \mathbb{R}^n$, $T : x_i \mapsto y_i$ is a transformation from $\mathbb{R}^n$ to $\mathbb{R}^m$, $\mathbf{\Phi}_l$ is an operator on the numerical sequence. We say $T$ is *weakly equivalent*, if for every $x$, which is a local minimal or maximals for $\mathbf{\Phi}_x(\{x_i\}_{i=0}^{N-1})$ then $\mathbf{\Phi}_x(\{T(x_i)\}_{i=0}^{N-1})$ is also a local minimal or maximal respectively.

A few easy observations and results follow. By definition strong equivalence implies weak equivalence. Moreover, we have the following propositions to determine weak equivalence.

*Proposition 1:* If $\mathbf{\Phi}_l$ is twice differentiable with respect to $l$, then $T$ is *weakly equivalent*, if for any $l$, where $\frac{\partial \mathbf{\Phi}_l(\{x_i\}_{i=0}^{N-1})}{\partial l} = 0$, the following conditions hold

$$\frac{\partial \mathbf{\Phi}_l(\{T(x_i)\}_{i=0}^{N-1})}{\partial l} = 0 \tag{15}$$

and

$$\frac{\partial^2 \mathbf{\Phi}_l(\{x_i\}_{i=0}^{N-1})}{\partial l^2} \cdot \frac{\partial^2 \mathbf{\Phi}_l(\{T(x_i)\}_{i=0}^{N-1})}{\partial l^2} \geq 0 \tag{16}$$

.

*Proof:* if $l$ is a local maximal or local minimal, then $\frac{\partial \mathbf{\Phi}_l(\{x_i\}_{i=0}^{N-1})}{\partial l} = 0$ and $\frac{\partial^2 \mathbf{\Phi}_l(\{x_i\}_{i=0}^{N-1})}{\partial l^2} \leq 0$ or $\frac{\partial^2 \mathbf{\Phi}_l(\{x_i\}_{i=0}^{N-1})}{\partial l^2} \geq 0$. By the definition of weak equivalence, (15) and (16) follow. ■

If $l \in \mathbb{Z}$, Then we have the following criterion to determine weak equivalence.

*Proposition 2:* $T$ is *weakly equivalent* for an operator $\mathbf{\Phi}_l$, where $l \in \mathbb{Z}$, if for any $l$, the following condition holds

$$(\mathbf{\Phi}_l(\{x_i\}_{i=0}^{N-1}) - \mathbf{\Phi}_{l-1}(\{x_i\}_{i=0}^{N-1})) \cdot (\mathbf{\Phi}_l(\{T(x_i)\}_{i=0}^{N-1}) - \mathbf{\Phi}_{l-1}(\{T(x_i)\}_{i=0}^{N-1})) \geq 0 \tag{17}$$

.

*Proof:* Without loss of generality, we assume $l$ is a local maximal for $\mathbf{\Phi}_l(\{x_i\}_{i=0}^{N-1})$. then $\mathbf{\Phi}_l(\{x_i\}_{i=0}^{N-1}) \geq \mathbf{\Phi}_{l-1}(\{x_i\}_{i=0}^{N-1})$ and $\mathbf{\Phi}_l(\{x_i\}_{i=0}^{N-1}) \geq \mathbf{\Phi}_{l+1}(\{x_i\}_{i=0}^{N-1})$. If (17) holds, we have $\mathbf{\Phi}_l(\{T(x_i)\}_{i=0}^{N-1}) \geq \mathbf{\Phi}_{l-1}(\{T(x_i)\}_{i=0}^{N-1})$ and $\mathbf{\Phi}_l(\{T(x_i)\}_{i=0}^{N-1}) \geq \mathbf{\Phi}_{l+1}(\{T(x_i)\}_{i=0}^{N-1})$. Thus $l$ is also a local maximal for $\mathbf{\Phi}_l(\{T(x_i)\}_{i=0}^{N-1})$. ■

As the importance of second-order statistics, specially we would like to investigate the weakly equivalent transformation for the correlation function. We first introduce a lemma.

*Lemma 1:* If the transformation $T$:$\mathbb{R}^n \rightarrow \mathbb{R}^n$ is an inner-product preserving isometry, i.e. $x^T y = T(x)^T T(y),\ \forall x, y \in \mathbb{R}^n$, then $T(x) = \mathbf{R}x$, where $\mathbf{R}$ is an orthogonal matrix. Hence T is a bijective isometry.

*Proof:* First let $x = y$, we have $||x||_2 = ||T(x)||_2$, i.e. $T$ preserves the Euclidean norm. Since $T(x_0)$ is on the ball $||x||_2 = ||x_0||_2$, we have $T(x) = \mathbf{R}(x)x$, where $\mathbf{R}(x)$ is an orthogonal matrix function. Let

$$\mathbf{R}(x) = \begin{pmatrix} u_1(x) & u_2(x) & \ldots & u(x) \end{pmatrix} \tag{18}$$

, where $\{u_i(x)\}$ is orthonormal. Furthermore, let $x = (1,0,0,\dots,0)^T$ and $y = (y_1,0,0,\dots 0)^T$. then we have

$$x^T y = T(x)^T T(y) \tag{19}$$

$$y_1 = x\mathbf{R}(x)^T \mathbf{R}(y) y = y_1 u_1^T((1,0,0,\dots,0)^T) u_1(y) \tag{20}$$

Therefore $u_1^T((1,0,0,\dots,0)^T)u_1(y) = 1$. By *Cauchy-Schwartz* Inequality, we have

$$1 = u_1^T((1,0,0,\dots,0)^T)u_1(y) \le ||u_1((1,0,0,\dots,0)^T)||_2 \cdot ||u_1(y)||_2 = 1 \tag{21}$$

The equality holds if and only if $u_1(y) = \lambda u_1((1,0,0,\dots,0)^T)$, but $||u_1(y)||_2 = 1$, thus $\lambda = 1$. Therefore
$u_1(y) = u_1((1,0,0,\dots,0)^T), \forall y \in \mathbb{R}^n$. By the same arguments we can show $u_i(y) = u_i(e_i)$, $\forall i = 1,\dots n$, where $e_i$ is the standard basis of $\mathbb{R}^n$. So $\mathbf{R}(x) = \mathbf{R}$ is a constant orthogonal matrix, thus $T(x) = \mathbf{R}x$. This also shows T is a bijective isometry. ■

For correlation function, we have the following theorem showing that generally speaking, rotation can be thought as the ”only” weakly equivalent transformation.

*Theorem 4:* For a fix length sequence, any transformation which only brings small enough changes to the inner product value under previous mapping will be a weakly equivalent transformation for correlation function. However, if the length goes to infinity, then rotation (or scaled rotation) is the only weakly equivalent transformation for correlation function.

*Proof:* Consider the vector sequence $\{x_i\}_{i=0}^{N-1}$. The correlation function is
$r_l(\{x_i\}_{i=0}^{N-1}.) = \frac{1}{N}\sum_{n=0}^{N-1} x^T(n)x(n-l)$. Then we have

$$r_l - r_{l-1} = \frac{1}{N}\sum_{n=0}^{N-1} x^T(n)(x(n-l) - x(n-l+1)) \tag{22}$$

After the transformation, we have the correlation function
$r'_l = r_l(T(\{x_i\}_{i=0}^{N-1})) = \frac{1}{N}\sum_{n=0}^{N-1} T(x(n))^T T(x(n-l))$ and $r'_l - r'_{l-1} = \frac{1}{N}\sum_{n=0}^{N-1} T(x(n))^T (T(x(n-l)) - T(x(n-l+1)))$. By Proposition 2. $T$ is weakly equivalent if $r'_l - r'_{l-1}$ has the same sign as $r_l - r_{l-1}$.

Consider the alphabet $\mathcal{A}' = \mathcal{A} \bigcup \{a-b|\, a,b \in \mathcal{A}\}$, which means we consider the symbol “$a-b$” as a new symbol and we extend the mapping $f$ on the newly added symbols as $f : (a-b) \mapsto (f(a) - f(b))$, which also extends the transformation $T$ for $f(a-b)$ be $T(f(a)) - T(f(b))$. Thus finding the weakly equivalent transformation is same to find the $T$ which preserve the sign at each $l$ of cross correlation function $R_l$ for the sequence $\{x_i\}_{i=0}^{N-1}$ and $\{x_i - x_{i+1}\}_{i=0}^{N-1}$.

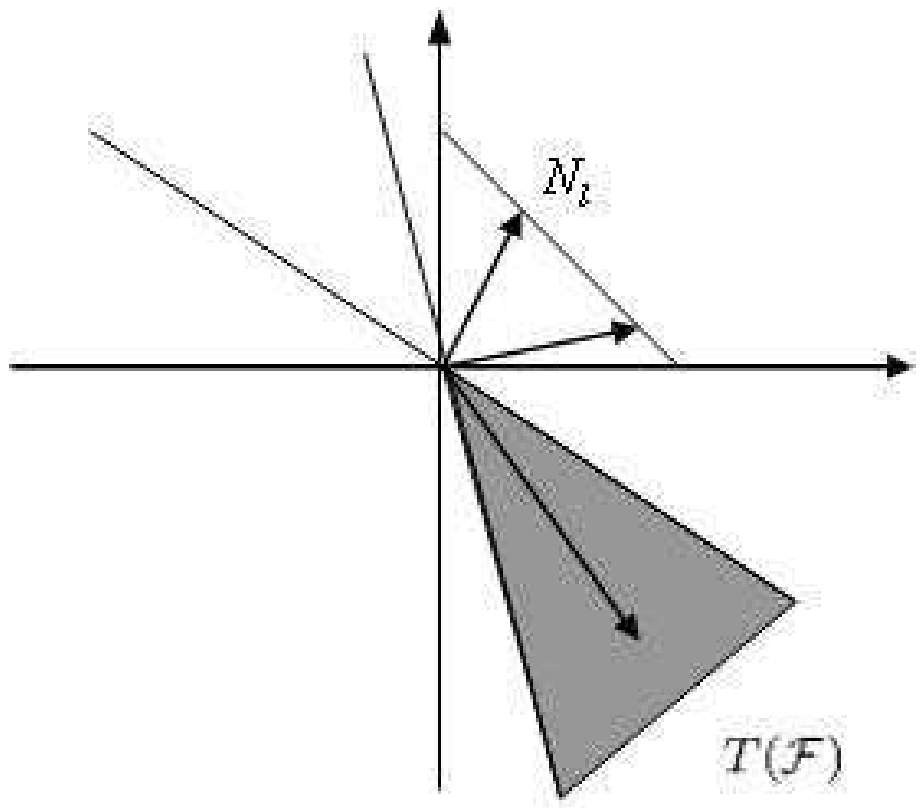

Figure 1. Illustration for $N_l$ and $T(\mathcal{F})$. $T(\mathcal{F})$ should reside in the convex cone as the shaded area in the figure.

Let $\mathcal{F} = \begin{pmatrix} a_1 & a_2 & \dots & a_A \end{pmatrix}^T$, where $a_i = f(x)^T f(y)$, $(x, y)\, or\, (y, x) \in \mathcal{A} \times \mathcal{A}'$ and $A = |\mathcal{A} \times \mathcal{A}'|$. Also let $N_l = \begin{pmatrix} C_1 & C_2 & \dots & C_A \end{pmatrix}^T$, where $C_i$ is the counting number for the pair $(x, y)$ corresponding to $a_i$ which appears in the cross correlation function $R_l$. Therefore we have

$$R_l = \frac{1}{N} \mathcal{F}^T N_l \tag{23}$$

Define $T(\mathcal{F}) = \begin{pmatrix} b_1 & b_2 & \dots & b_A \end{pmatrix}^T$, where $b_i = T(f(x))^T T(f(y))$, where $(x, y)$ corresponds to $(x, y)$in the $a_i$ in $\mathcal{F}$. Notice that $N_l$ will not change since it is determined by the given sequence. After the transformation, the cross correlation functions becomes $R'_l = \frac{1}{N} T(\mathcal{F})^T N_l$. We need $R_l$ and $R'_l$ have the same sign for all $l$. Notice that $\sum_{i=1}^{A} C_i = N$. So every $N_l$ for a given sequence corresponds to a point on the hyperplane $\sum_{i=1}^{A} x_i = N$ in $\mathbb{R}^A$. If $T$ preservers signs for all $l$, then for each $N_l$, $T(\mathcal{F})$ should reside in the same half plane of $\mathcal{F}$. Because $T(\mathcal{F})^T N_l$ should have the same sign of $\mathcal{F}^T N_l$. In general, the sign will not all be positive or negative, since that will means the correlation function is monotonic which in general is not valid for all sequences. Consider all possible symbol sequence of length $N$ and $l \in \mathbb{Z}$. Then

$$T(\mathcal{F}) \in \bigcap_{all\, sequences\, of\, length\, N} \{x \mid (x^T N_l) \cdot (\mathcal{F}^T N_l) \geq 0\} \tag{24}$$

This implies $T$ should reside in the intersection of all half plane determined by all sequences of length $N$. Each half plane is a convex cone, therefore the intersection is still a convex cone as illustrated in Fig. 1. Since we have finitely many point on the hyperplane $\sum_{i=1}^{A} x_i = N$ . All $N_l$ reside in the first quadrant. We can always construct two hyperplanes whose intersection is in one quadrant. Thus the intersection will be a small convex cone in a quadrant. If $T(\mathcal{F})$ is in that convex cone, then the sign is all preserved, which means $T$ is a weakly equivalent transformation. This proves the first claim.

However, if we let $N$ go to infinity, first notice that the intersection will not be empty set, since $\mathcal{F}$ is always in the intersection. But the points $N_l$ become dense in the first quadrant. We can have a sequence of $N_l$ such that $N_l^T \mathcal{F} \to 0^+$ and $N_l'$ with $N_l' \mathcal{F} \to 0^-$. Therefore the intersection will be squeezed to the line $\lambda\mathcal{F},\ \lambda \geq 0$. i.e. The $T(\mathcal{F}) = \lambda\mathcal{F}$. If the scaling $\lambda = 1$, then $T(f(x))^T T(f(y)) = f(x)^T f(y),\ \forall f$. Since the mapping is arbitrary, which means it should hold for any $x$ and $y$. By lemma 1, $T$ is a rotation or a scaled rotation.

Conversely, by theorem 1, we have rotation or scaled rotation is a strongly equivalent transformation. Thus it follows it is also a weakly equivalent transformation. ■

From this theorem, we can see that rotation is essentially the only weakly equivalent transformation for correlation function. We can expand the class of operator having this property by combining theorem 4 and theorem 3 with some technical conditions, then we have the following corollary.

*Corollary 1:* If a second order operator whose Hessian is of the form as (10) and all $k_i$ have the same sign and $0 < \sum k_i < \infty$, then rotation is essentially the only transformation which is both strongly and weakly equivalent.

*Proof:* If the Hessian has the form above, then

$$\mathbf{\Phi}_l = \frac{1}{2} sgn(k_1) \sum_{n=0}^{N-1} |k_n| x(n)^T x(n-l) \tag{25}$$

, Where

$$sgn(x) = \begin{cases} 1 & x \geq 0 \\ -1 & x < 0 \end{cases}$$

Notice that the proof for correlation function follows here except we shall use $N'_l = \begin{pmatrix} \zeta_1 C_1 & \zeta_2 C_2 & \dots & \zeta_A C_A \end{pmatrix}^T$ instead of $N_l$, where $\zeta_i$ are all positive and

$$\zeta_i = \sum_{j|all\ terms\ having\ same\ pattern} k_j$$

If the length goes to infinity, $N_l$ becomes dense, since $0 < \sum k_i < \infty$, then $N'_l$ is in some non-degenerated bounded set, which is still dense and resides in the first quadrant. Then the argument above still valid. Thus rotation is essentially the only weakly equivalent for this kind of operator. By theorem 1, rotation is also the strongly equivalent transformation. ■

### *C. Mapping Between 1D Euclidean Spaces*

The case in which the transformation $T : \mathbb{R}^n \to \mathbb{R}^n$ is limited to $n = 1$ is particularly common in the literature. Moreover, this case stands out and deserves special attention since the only possible rotation on $\mathbb{R}$ is obtained by scaling and interchanging the mapping values. From Theorems 1 and 4, we observe that if the mappings cannot be obtained by scaling and interchanging the mapping values, the correlation and Fourier analysis results obtained using these mappings are neither strongly nor weakly equivalent. Therefore, in the 1D case, the equivalent mapping class under a given operator becomes fairly limited. In general, distinct mappings will usually lead to inconsistent correlation and Fourier analysis results. Another interesting fact about 1D mappings that can be derived from our previous results is that if the mapping is binary (i.e. the range of the mapping can only takes two distinct values), then we observe that the correlation and Fourier analysis under any two such binary mappings are always consistent since we can always obtain one of the mappings by scaling and interchanging the mapping values of the other mapping.

### *D. Mapping Between Euclidean Spaces of Different Dimensions*

In the previous sections, we focused primarily on the transformation $T : \mathbb{R}^n \to \mathbb{R}^m$, where $n = m$. In this section, we will present a brief discussion of the case where $m \neq n$. If $m > n$, which means $T$ will transform the vector into a larger dimensional Euclidean space. However, since there is a natural embedding for $\mathbb{R}^n$ into $\mathbb{R}^m$, we can always think the transformation as $T'$:$\mathbb{R}^m \to \mathbb{R}^m$. For second-order operators which are shown equivalent under rotation, we still have the same results in this situation, except the rotation matrix here means a matrix have orthonormal columns.

For the case $m < n$, we can also think as $\mathbb{R}^m$ is embedded inside $\mathbb{R}^n$ by the transform

$$y = \begin{pmatrix} I_{m\times m} \\ 0_{(n-m)\times m} \end{pmatrix}_{n\times m} x \tag{26}$$

, where $x \in \mathbb{R}^m, y \in \mathbb{R}^n$. Then we only need to research on the new transformation $T'$:$\mathbb{R}^n \to \mathbb{R}^n$. However, in this case, we can see that we actually project the higher dimensional subspace into a lower dimensional space, rotation here in general is not an equivalent transformation anymore. Intuitively, because of the projection, we lose information projected on $(n-m)$ dimensions. Therefore rotation is no longer an equivalent transformation.

## III. Abstract Mapping Equivalence for Symbolic Sequences

### *A. Abstract Mapping Model and Examples*

In previous sections, we mainly focused on properties of mappings, which map symbols into vector space. However, it is not necessary to restrict to the vector space. Many classical concepts in numerical signal processing can be extended to various algebraic structures. For example, the Fourier transform and Wavelet transform can be defined on group, ring and finite-field [33], [34], [35]. In this section, we introduce the generalized mapping to arbitrary semi-ring, ring or algebra structure. We shall also extend the notion of equivalence defined in previous section.

For given finite alphabet $\mathcal{A}$, we define $F(\mathcal{A})$ as collection of all the symbolic sequences and define the binary operation as concatenating two symbolic sequences. It can be shown $F(\mathcal{A})$ is a free semi-group [36]. $R$ is any semi-ring. Let $\mathcal{R}$ to be collection of all maps from $F(\mathcal{A})$ to $R$. For any $f \in \mathcal{R}$, we denote it as the formal series,

$$f = \sum_{u\in F(\mathcal{A})} f(u)u \tag{27}$$

we define two operation $+$ and $\cdot$ on $\mathcal{R}$ as

$$f + g = \sum_{u\in F(\mathcal{A})} (f(u) + g(u))u \tag{28}$$

$$(f \cdot g)(s) = \sum_{uv=s} (f(u)g(v))s \tag{29}$$

With these two binary operations, we have the following proposition to show that we construct a new algebraic structure on $\mathcal{R}$

*Proposition 3:* $R$ is a semi-ring or ring, then $(\mathcal{R}, +, \cdot)$ forms a semi-ring or ring respectively.

*Proof:* By the definition of addition, we can see that if $(R, +)$ is a commutative monoid or abelian group, then $(\mathcal{R}, +)$ has the same property correspondingly. Therefore it's enough to show to show $(\mathcal{R}, \cdot)$ is a semi-group, i.e. we need to show the multiplication is associative.

$\forall f, g, h \in \mathcal{R}$

$$\begin{aligned}(fg)h(s) &= \sum_{xw=s} (fg(x))h(w) \\ &= \sum_{xw=s} \sum_{uv=x} (f(u)g(v))h(w) \\ &= \sum_{uvw=s} (f(u)g(v))h(w) \\ &= \sum_{u(vw)=s} f(u)(g(v)h(w)) \\ &= f(gh)(s) \end{aligned} \tag{30}$$

Therefore the multiplication is associative. We proved the proposition. ■

The ring $\mathcal{R}$ is called as the semi-group ring of $F(\mathcal{A})$ with coefficients in $R$. Furthermore, if $R$ is a left-$R'$ module for some ring $R'$. We can define for any $r \in R'$,

$$rf = \sum_{u \in F(\mathcal{A})} rf(u)u \tag{31}$$

then $\mathcal{R}$ is the left-$R'$ algebra. The $\mathcal{R}$ may be interpreted as the generalized filter space, while the $F(\mathcal{A})$ is as the signal space. The multiplication can been thought as the extension of discrete convolution. If we let $R$ and $R'$ to be $\mathbb{R}$, $\mathcal{A} = \{0, 1, 2, \dots\}$, the multiplication degenerates to classical convolution. The symbol sequence is mapped into a numerical sequence.

Another example of the abstract mapping model is the probability model. Consider all the outcomes of the words in $F(\mathcal{A})$. Denote the outcome space as $\Omega$. $\omega$ is a $\sigma$-algebra on $\Omega$ and $\mathbb{P}$ is a probability measure on $\sigma$. Notice that two set-operations on $\sigma$, $\cap$ and $\cup$ is analog of $\cdot$ and $+$. $\emptyset$ and $\Omega$ can be seen as 0 and 1 respectively. Therefore $R = (\omega, \cup, \cap)$ forms a semi-ring. The probability measure mapping $P$ here is interpreted as a semi-ring mapping from $\mathcal{R}$ to the semi-group ring of $F(\mathcal{A})$ with coefficients in $\mathbb{R}$, which is defined as,

$$P(f) = \sum_{u \in F(\mathcal{A})} \mathbb{P}(f(u))u \tag{32}$$

The probability operations then can be realized by algebraic operations on $\mathcal{R}$ and the corresponding probability measure values are obtained after the mapping $P$.

### *B. Abstract Mapping Equivalence*

For generalized mapping, the equivalence problem is still worth for investigating. However, in the situation, it becomes much more difficult than in a $\mathbb{R}$-vector space. The $R$ in general does not possess any meaningful ordering. Therefore the definition of equivalence turns out to be limited for specific application. Nevertheless, as we mentioned before, in most cases, it is reasonable to require the result to be similar in certain extent. From now on, we assume $R$ is a integral domain with unity 1. We introduce the following definition for abstract equivalence of a generalized mapping.

*Definition 4:* For any $f, g \in \mathcal{R}$, $f$ and $g$ are abstractly equivalent, if the ideals they generated are the same, i.e. $(f) = (g)$.

The next proposition shows the intuition and legitimacy of this definition.

*Proposition 4:* $(f) = (g)$ if and only if $f = ug$, where u has multiplicative inverse.

*Proof:* If $(f) = (g)$, then $f = u_1 g$ and $g = u_2 f$ for some $u_1, u_2 \in \mathcal{R}$. We have

$$\begin{aligned} f &= u_1 u_2 f \\ (1 - u_1 u_2) f &= 0 \end{aligned} \tag{33}$$

Since $R$ is integral domain, we have $u_1 u_2 = 1$. $u_1$ and $u_2$ are units.

Conversely, if $f = ug$, then $g = u^{-1} f$. We have $(f) \subseteq (g)$ and $(g) \subseteq (f)$, therefore $(f) = (g)$. ■

A loose interpretation of Proposition 4 implies that abstractly equivalent mappings only differ by a "scale" and that the scale change can be "reversed." Let us first consider the case where the semi-ring $R$ is $\mathbb{R}$ or $\mathbb{C}$. In this case, $R$ forms a field and thus any non-zero element is a unit. It is easy to show that in this case strong equivalence implies abstract equivalence. To see that strong equivalence is a special case of abstract equivalence, let us consider mappings $f$ and $g$ to be defined at the origin "0" of the field (i.e. we ignore the translation between the mappings). If non-trivial mappings $f$ and $g$ are strongly equivalent, then $f = cg$, where $c$ is a non-zero real or complex number. We observe that $c$ is a unit and therefore its inverse $c^{-1}$ exists. Finally, we note that the strongly equivalent mappings $f$ and $g$ are abstractly equivalent.

We now extend the discussion to semi-ring $R$ given by $\mathbb{R}^n$ or $\mathbb{C}^n$. We note that $R$ forms a vector space over $\mathbb{R}$ or $\mathbb{C}$. We recall that the orthogonal linear operator is a necessary and sufficient condition for strong equivalence under the correlation function. Moreover, we note that the set of orthogonal linear operators forms an orthogonal group $O(n)$ given by $O(n) = \{M \in \mathbb{C}^{n\times n} : M^H M = I\}$, where $I$ denotes the identity operator. The orthogonal group contains the special orthogonal group $SO(n)$ which represents usual rotations and is given by $SO(n) = \{M \in \mathbb{C}^{n\times n} : M^H M = I \; and \; det(M) = 1\}$. Finally, we observe that if $f$ and $g$ are strongly equivalent under the correlation function, then $f = Mg$, where $M \in O(n)$. Since $M$ is in the orthogonal group $O(n)$, we note that it is a unit (i.e. $M^{-1} = M^H \in O(n)$). Therefore, we once again conclude that strong equivalence implies abstract equivalence.

## IV. Applications and Examples in Genomic Signal Processing

As we discussed in previous sections, approach of mapping the symbolic sequence to $\mathbb{R}^n$ is a widely adopted method for symbolic signal process. Therefore the consistency problem for results using different mappings always arises. In this section, we will apply our theory to genomic signal processing.

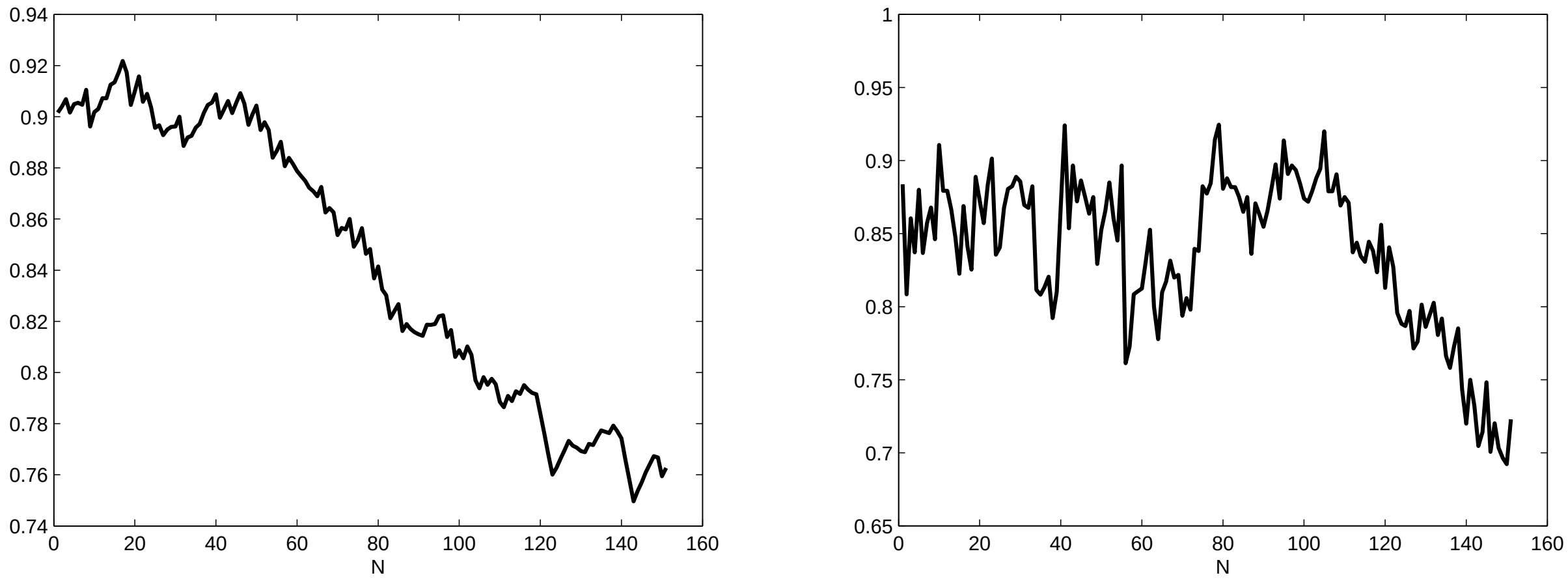

(a) Correlation coefficients for strong equivalence

(b) Percentage of points preserving local extremes for weak equivalence

Figure 2. Two consistency measurements for correlation results using two mapping methods change with the growth of sequence length $N$ for AD169 DNA sequence.

We conduct experiments on Human gene AD169 sequence (GenBank accession no. X17403). We calculate the correlation function as in (6) using two different mappings. The first one maps the $\mathcal{A} = \{\texttt{A},\ \texttt{T},\ \texttt{G},\ \texttt{C}\}$ to the standard basis of $\mathbb{R}^4$ correspondingly. Then we use another mapping strategy, which maps A to $(-1, 0, 0, 0)$, T to $(1, 0, 0, 0)$, G to $(0, 1, 0, 0)$ and C to $(0, -1, 0, 0)$. These are two widely used mapping methods [11], [12]. In Fig. 2(a), we show the changing of correlation coefficient between the two correlation results with growth of DNA sequence length $N$ and in (b) we show how the percentage of the points having same local extremum property in two results grows with $N$. The second mapping is not obtained by rotation of the first mapping. As a result, all these two metrics have a decreasing trend with the grown of length $N$.

We also calculate the two metrics on rhodopsin gene sequence (GenBank accession no. U49742) for these two metrics and the result is shown in Fig. 3.

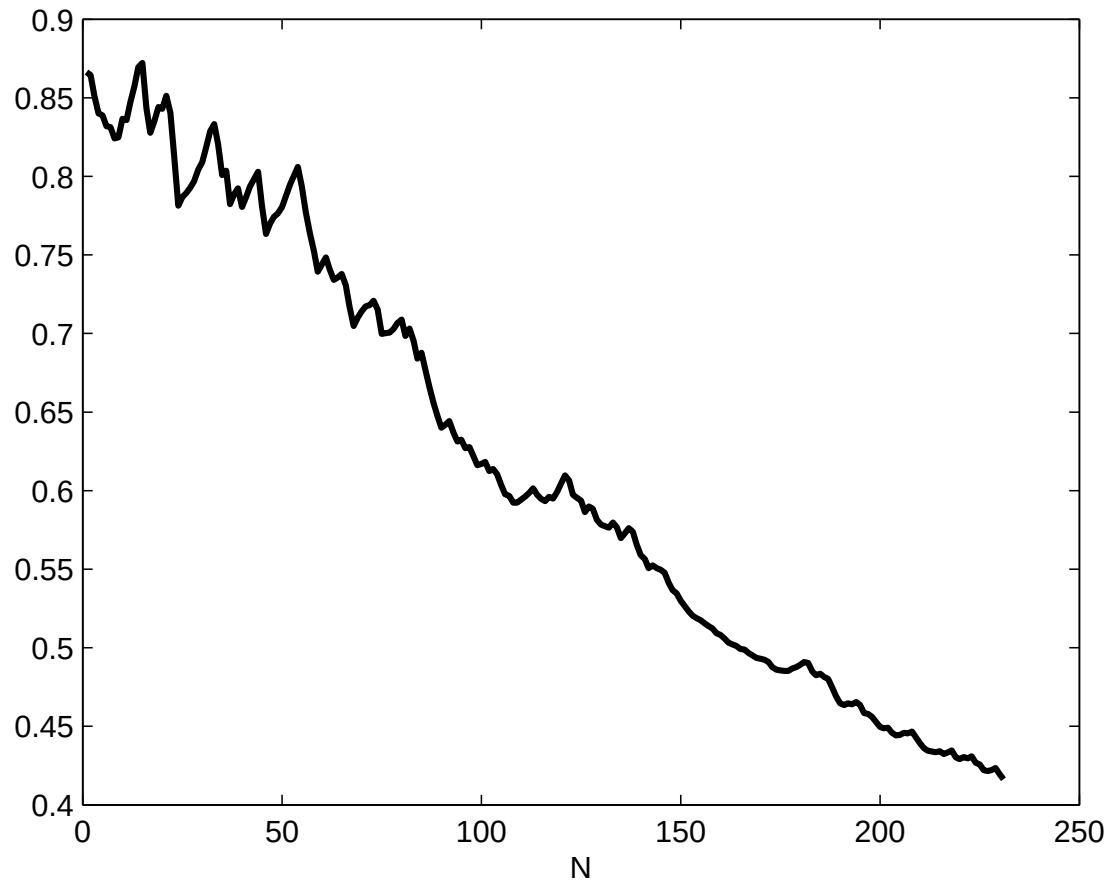

(a) Correlation coefficients for strong equivalence

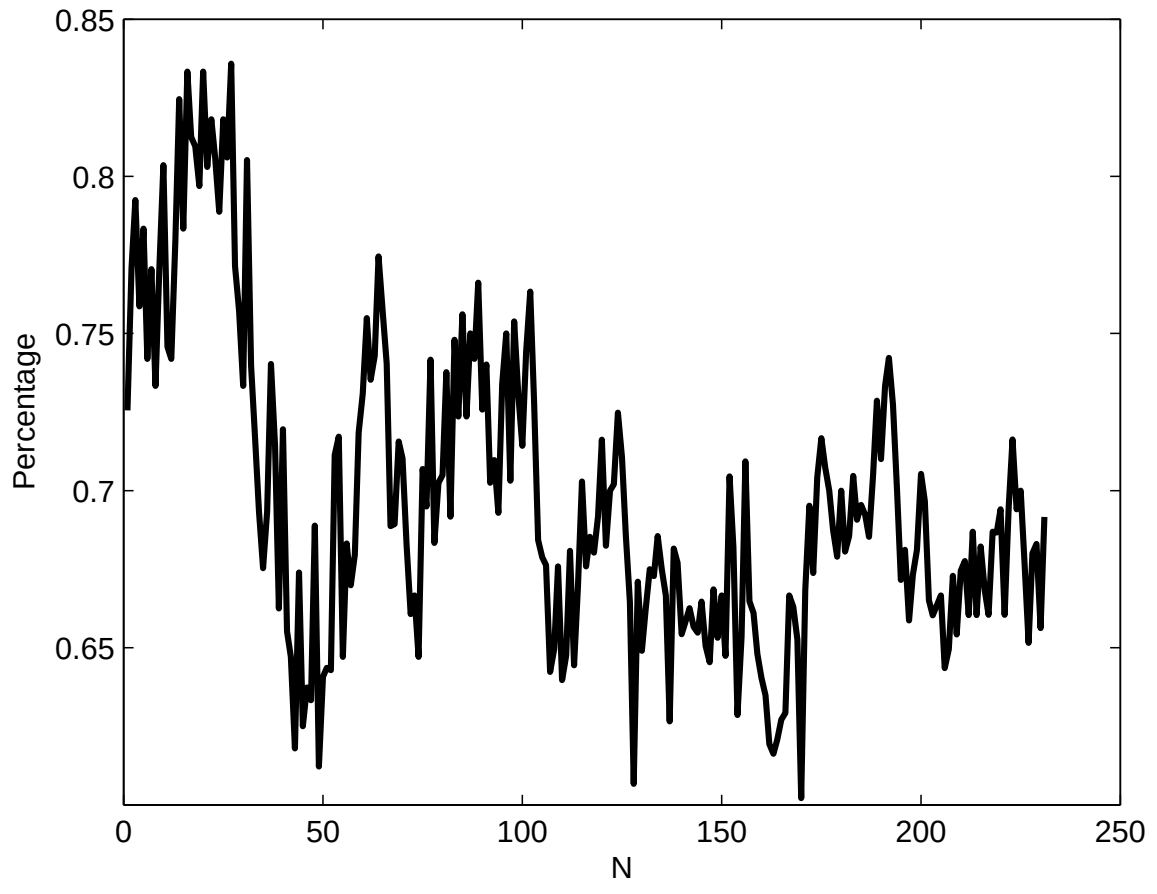

(b) Percentage of points preserving local extremes for weak equivalence

Figure 3. Two consistency measurements for correlation results using two mapping methods change with the growth of sequence length $N$ for rhodopsin gene sequence.

These two examples show the same trends for two metrics between two mappings. The similarity between the two results become less and less, which finally may lead to an inconsistent analysis results due to the fact that two chosen mapping methods are not equivalent for the correlation function. Thus it does not make sense to make comparison between the analysis result for a given gene sequence under these two mapping methods.

In Fig. 5, we show the two consistency measurements between the power spectrum under the previous two mapping methods. Figures 5(a) and 5(b) show the correlation coefficient and the percentage of the

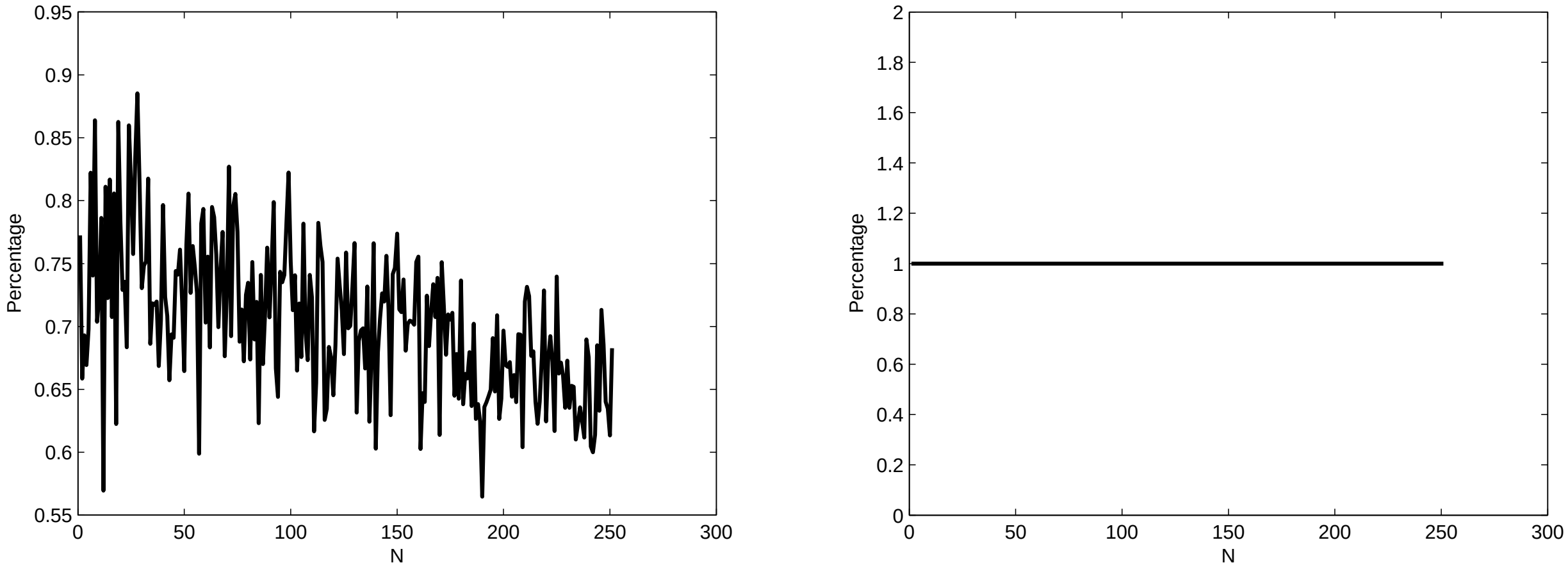

(a) Weak equivalence metric between results using the first and the second mapping

(b) Weak equivalence metric between results using the first and the third mapping

Figure 4. Percentage of points preserving local extremes for Fourier transform using three different maps changes with growth of sequence length $N$ for human gene AD169 sequences.

points having same local extremum property in two power spectrum results grow with the sequence length $N$ for human gene AD169 respectively. (c) and (d) show the strong and weak equivalence measurements changing with the length $N$ respectively for rhodopsin gene sequences. Although the equivalent transform we analyzed before does not mainly focus on power spectrum, we can still find that the power spectrum results using these two different mappings have the trend to be inconsistent. Since the correlation and power spectrum are widely used and pervasive in statistic analysis, it suggests the consistency problem should not be neglected when comparing analysis results.

Research on statistical properties of coding and non-coding regions in nucleotide sequences is an important topic in genomic signal processing [27], [28]. We shall also conduct experiments on coding and non-coding regions of Human gene TXNDC9 and NOC2L using the two mapping methods introduced earlier. As shown in Fig. 6, for both genes, the consistency measure between the correlation functions decays as the length $N$ increases. As a result of our analysis, we note that the correlation results under the two mappings are inconsistent in the long run. Furthermore, any comparison between the analysis results obtained by relying on these correlation functions becomes increasingly unreliable. Another interesting result can be observed in Fig. 6, where the decay rate of the non-coding region is faster than the coding region. This phenomenon could be attributed to the fact that the coding regions can be viewed as more

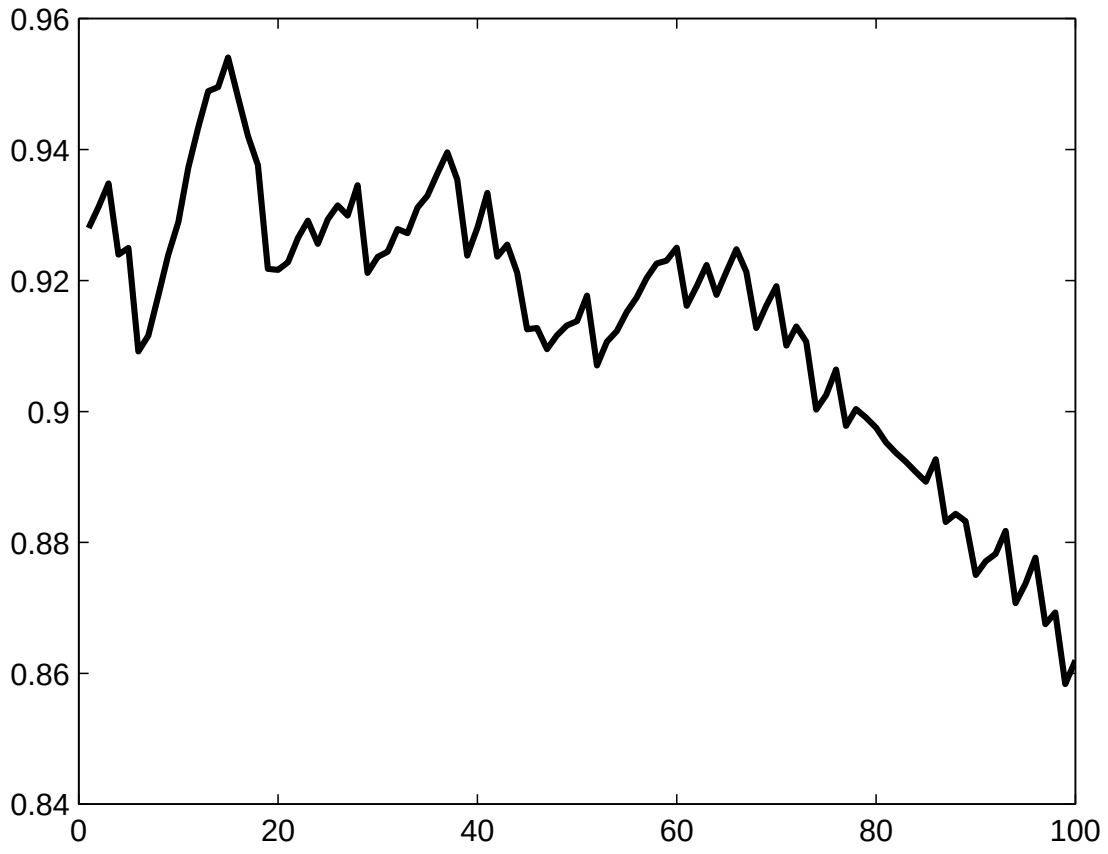

(a) Correlation coefficients for strong equivalence

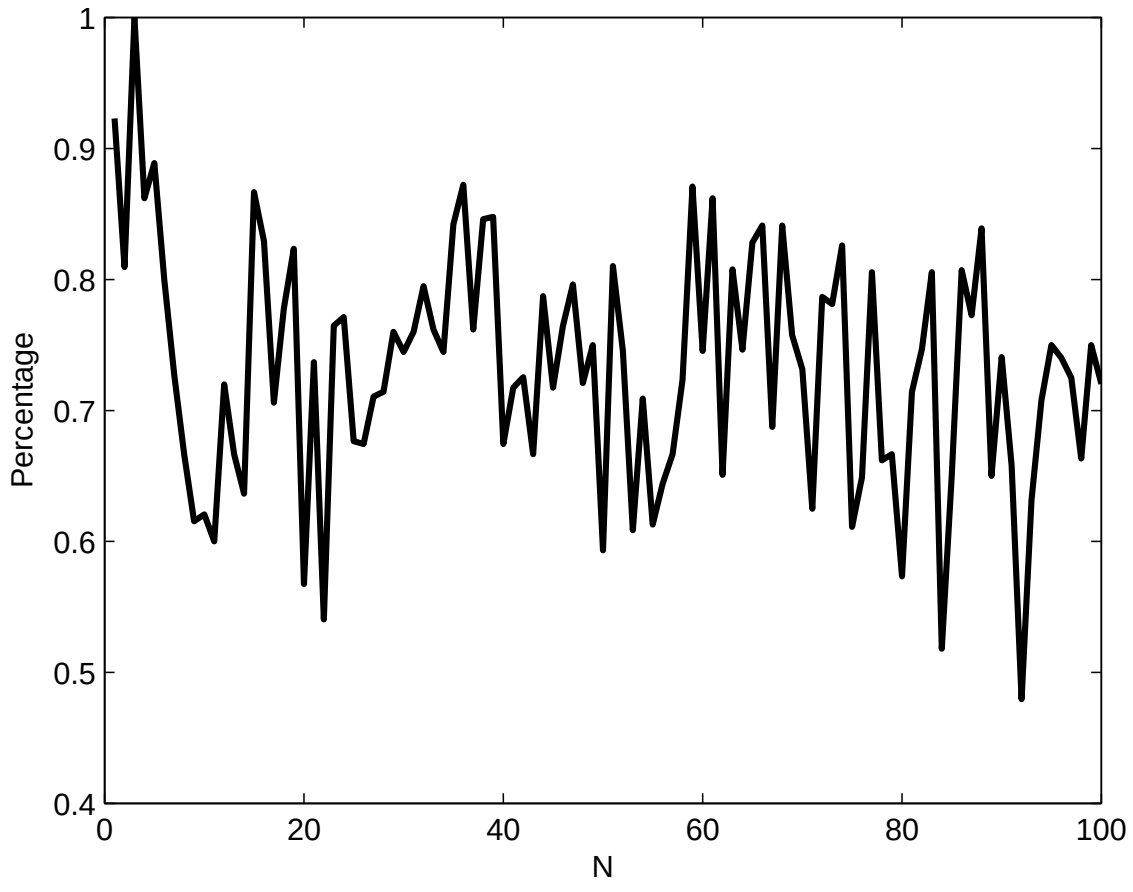

(b) Percentage of points preserving local extremes for weak equivalence

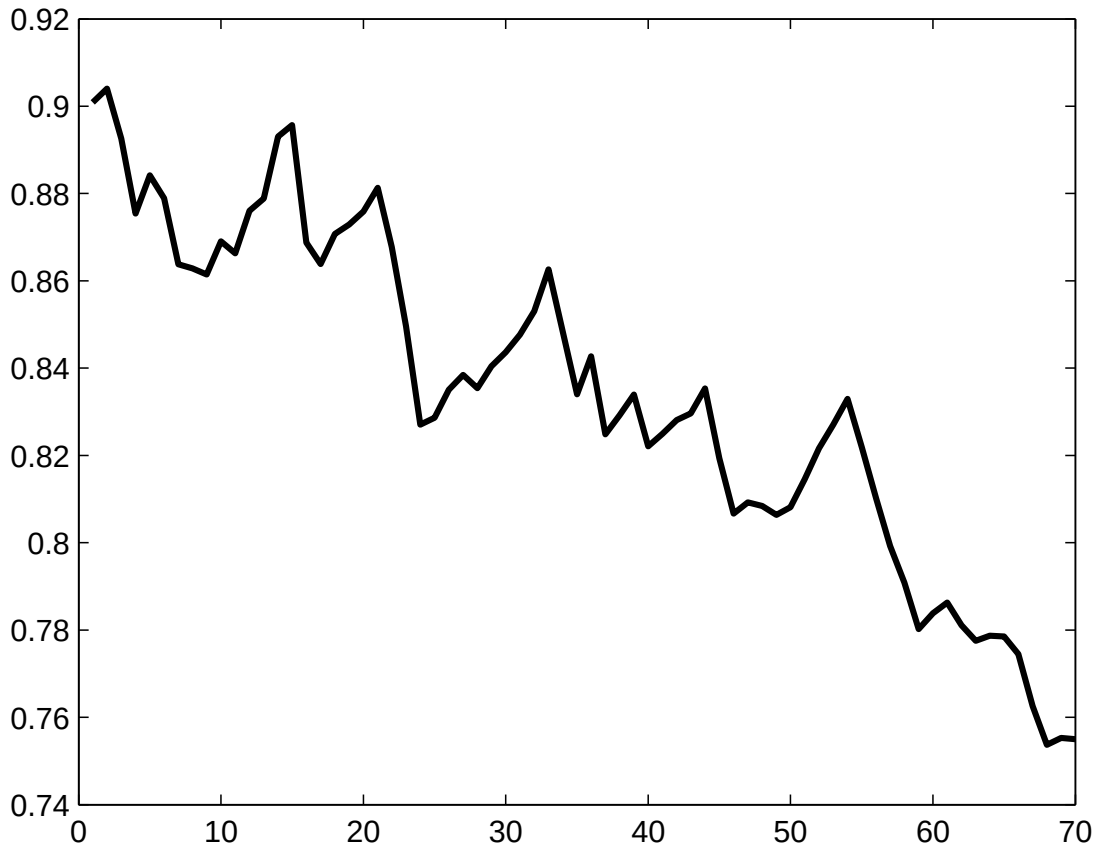

(c) Correlation coefficients for strong equivalence

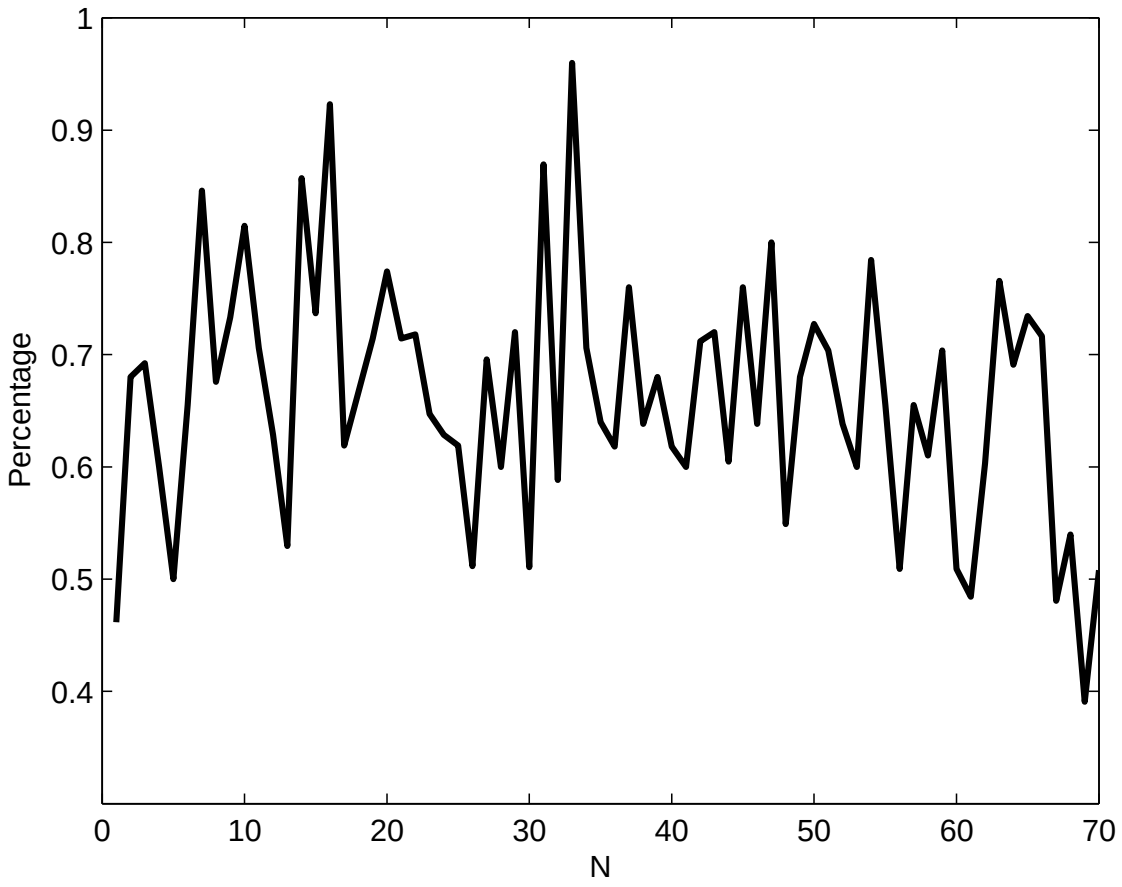

(d) Percentage of points preserving local extremes for weak equivalence

Figure 5. (a), (b) shows the correlation coefficient and percentage of points preserving local extremes of the power spectrum change with growth of sequence length $N$ for Human gene AD169 sequences respectively using mapping which maps $\mathcal{A} = \{\texttt{A},\ \texttt{T},\ \texttt{G},\ \texttt{C}\}$ to the standard basis of $\mathbb{R}^4$. (c) and (d) shows same two consistency measurements respectively change with growth of sequence length $N$ for rhodopsin gene sequences using the map which maps $\texttt{A}$ to $(-1, 0, 0, 0)$, $\texttt{T}$ to $(1, 0, 0, 0)$, $\texttt{G}$ to $(0, 1, 0, 0)$ and $\texttt{C}$ to $(0, -1, 0, 0)$.

random than the non-coding regions. Nevertheless, the main conclusion that we draw our attention to is that the consistency of the correlation between non-equivalent mappings decays as the sequence length increases for both coding and non-coding regions.

We also calculate the Fourier transform as defined in (11) on Human gene AD169 sequence. The first mapping is chosen as before, which maps the $\mathcal{A} = \{\texttt{A},\ \texttt{T},\ \texttt{G},\ \texttt{C}\}$ to the standard basis of

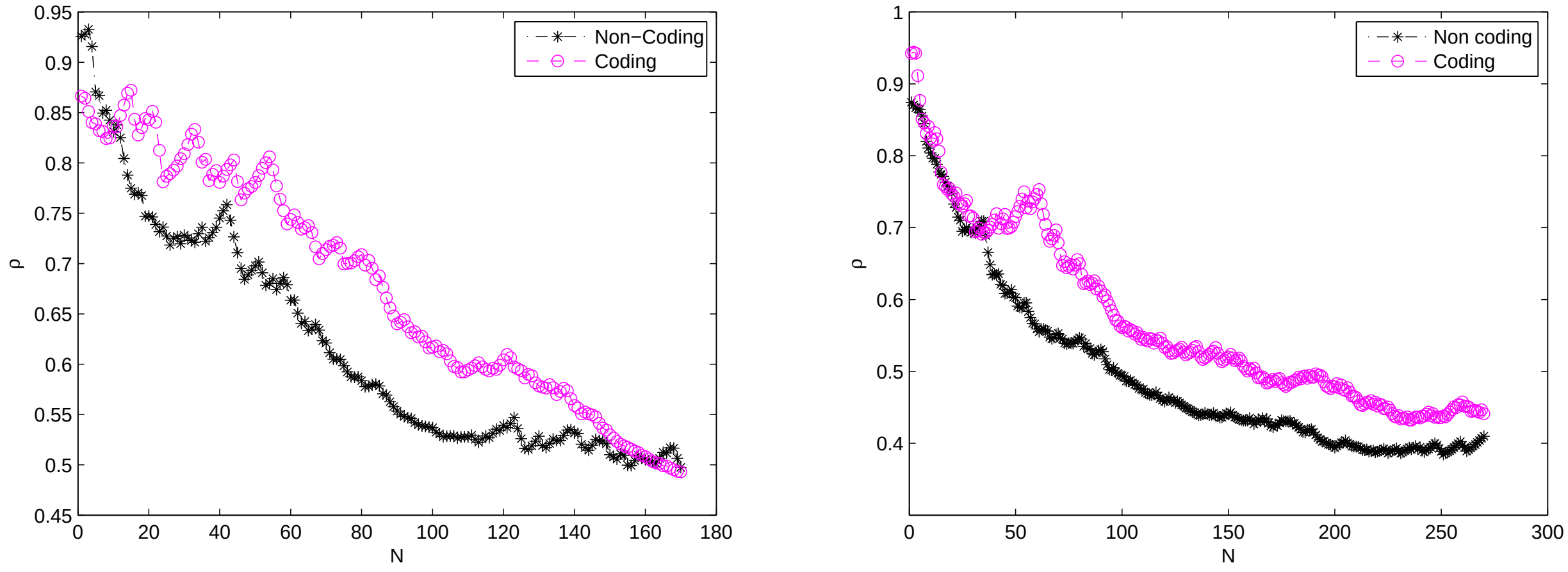

(a) Correlation coefficients for strong equivalence of coding and non-coding regions of TXNDC9

(b) Correlation coefficients for strong equivalence of coding and non-coding regions of NOC2L

Figure 6. Comparison the consistency measure of correlation functions changing with growth of $N$ for coding and non-coding regions of Human gene TXNDC9 and NOC2L.

$\mathbb{R}^4$ respectively. Then we use second mapping strategy, which maps `A` to $(0.9912, 0.1322, 0, 0)$, `T` to $(0.8367, -0.239, 0.1195, 0.4781)$, `G` to $(-0.7505, -0.5361, -0.2144, 0.3216)$ and `C` to $(0.7804, -0.5103, -0.2401, -0.2701)$. The third strategy maps `A` to $(\frac{1}{\sqrt{2}}, 0, \frac{1}{\sqrt{2}}, 0)$, `T` to $(0, \frac{1}{\sqrt{2}}, 0, \frac{1}{\sqrt{2}})$, `G` to $(-\frac{1}{\sqrt{2}}, 0, \frac{1}{\sqrt{2}}, 0)$ and `C` to $(0, -\frac{1}{\sqrt{2}}, 0, \frac{1}{\sqrt{2}})$. We have normalized the mappings so that it will not change the energy of the result. The second mapping is not obtained by rotating the first mapping. While the third mapping is obtained by rotating the first mapping. In Fig. 4, we show the weak equivalence metrics for Fourier analysis results. Figure 4(a) shows consistency between 1 and 2 results becoming less and less. While (b) suggests a completely consistent results. In Fig. 7, we show the analysis result of the three mappings. As we showed before, (a) and (c) are exactly same, since rotation is a strongly equivalent transformation. We can find many differences between (a) and (b), especially at the peaks. We also calculate the correlation coefficient between (a) and (b), which is $0.82$. The peak here means the repeat pattern of some periodic sequences, however, since we have shown that the mapping is not equivalent here, it makes no reason to debate on possible conflicting analysis results for this gene sequence.

In Fig. 8, we illustrate consistency measures for the mapping $T : \mathbb{R}^n \to \mathbb{R}^m$, where $m \neq n$. The first mapping we used is the standard Voss mapping introduced earlier. The second mapping we employed is the RY rule [9], which maps `A`, `G` to $1$ and `T`, `C` to $-1$. In this case, the transformation $T : \mathbb{R}^4 \to \mathbb{R}^1$ is not

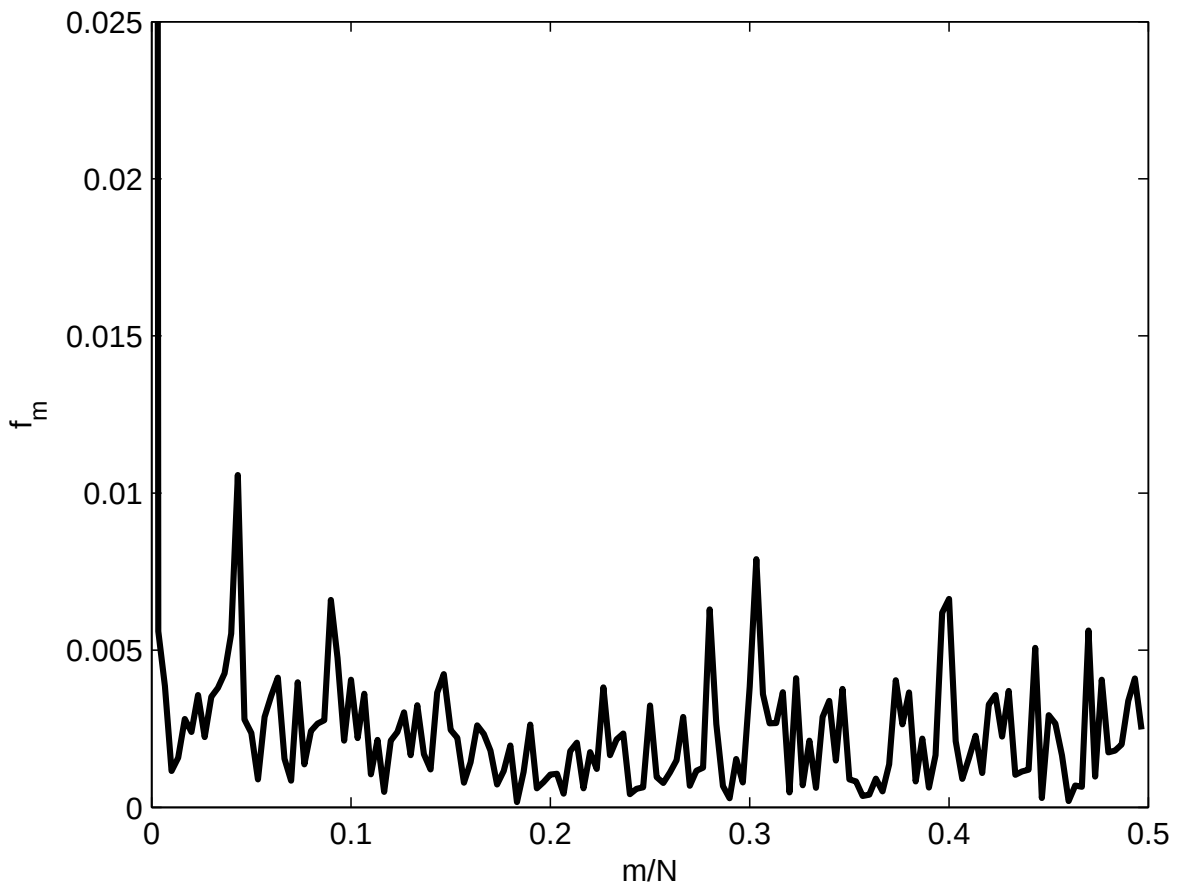

(a) Fourier transform under the first mapping

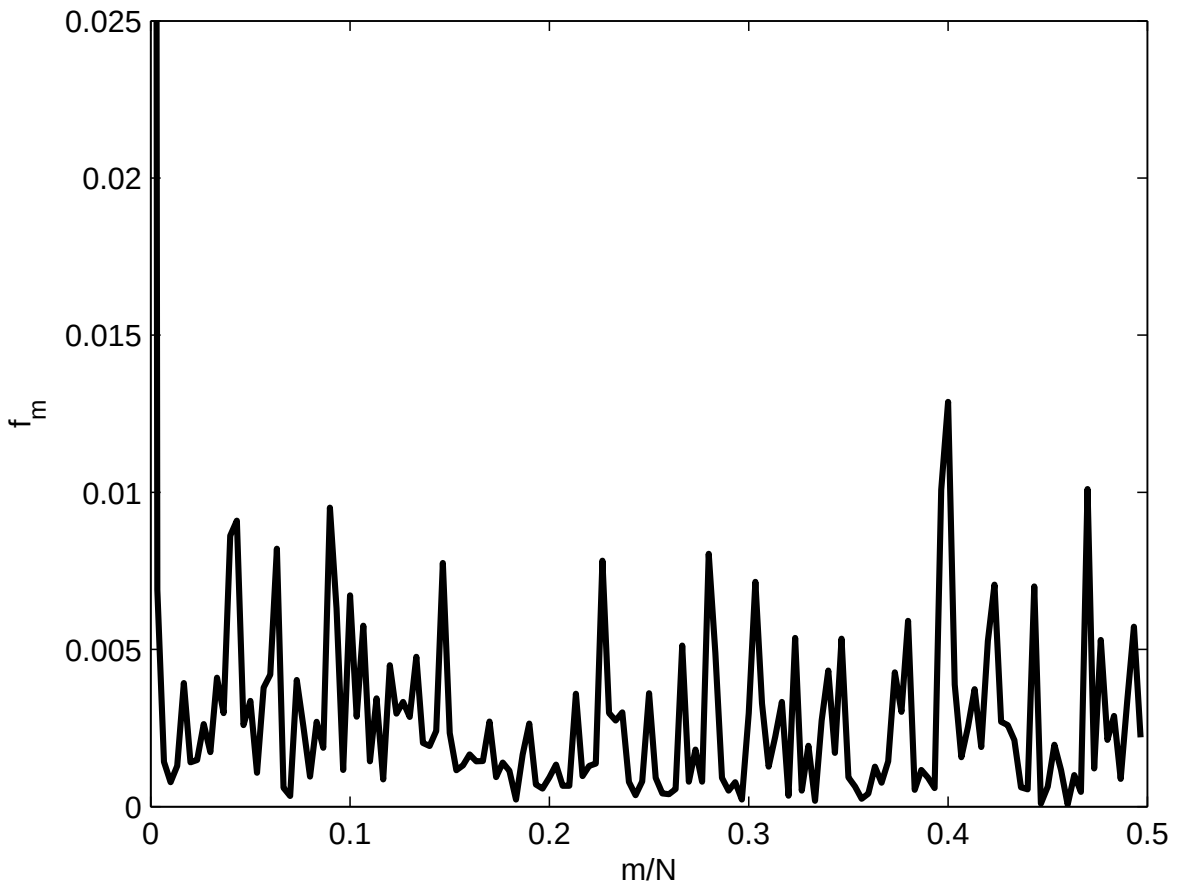

(b) Fourier transform under the second mapping

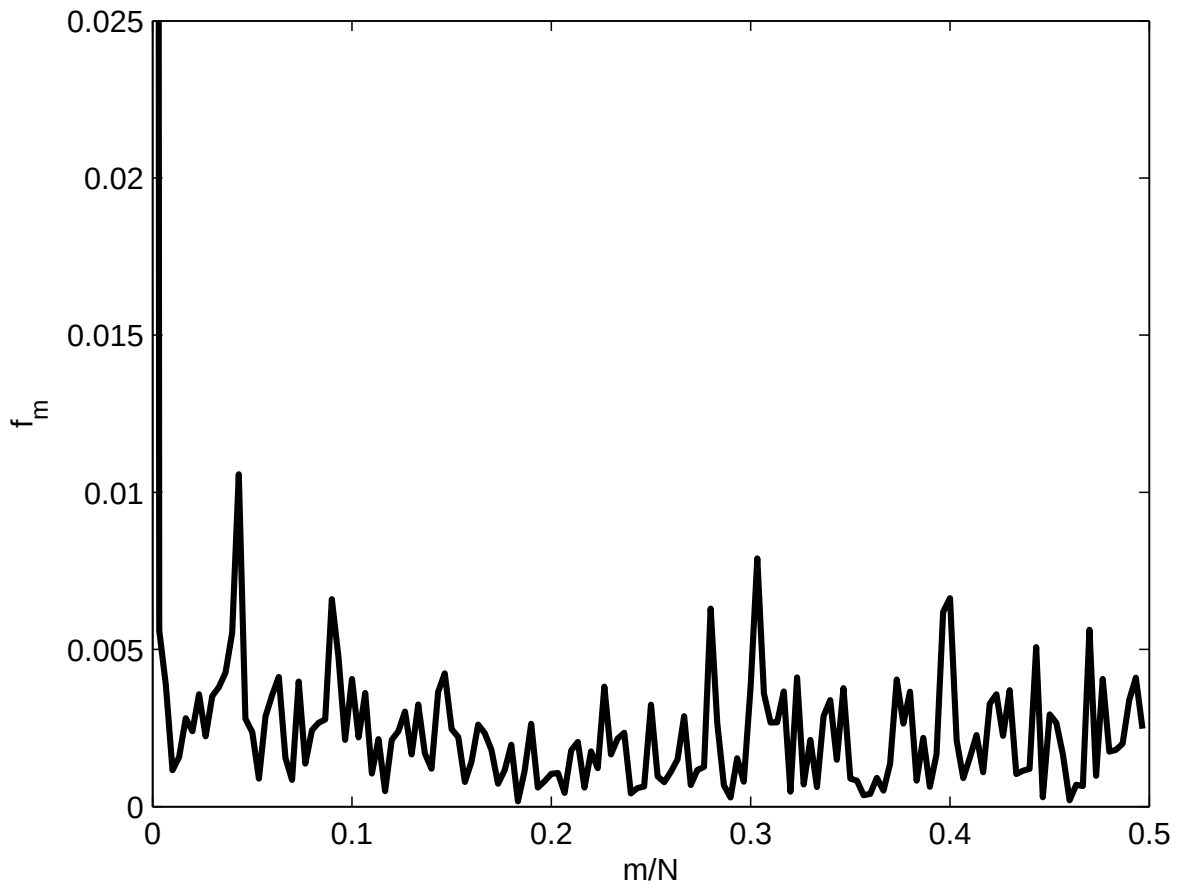

(c) Fourier transform under the third mapping

Figure 7. Fourier transform using three mapping methods.

an equivalent transformation as discussed in Section II-D. From Fig. 8, we observe that the consistency between the mappings decays as sequence length $N$ increases.

In all of the experiments conducted we observe that rotation serves as the unique equivalent transformation for the correlation function. Rotation also provides a strongly equivalent transformation for Fourier and spectrum analysis. Mappings which are not equivalent lead to inconsistent results as the sequence length $N$ increases. However, we must point out that the opposite may not be true: specifically, for a fixed-length sequence, the consistency for any two mappings does not necessarily decay as the difference between these two mappings increases, measured in the sense of rotation equivalence, i.e. the similarity

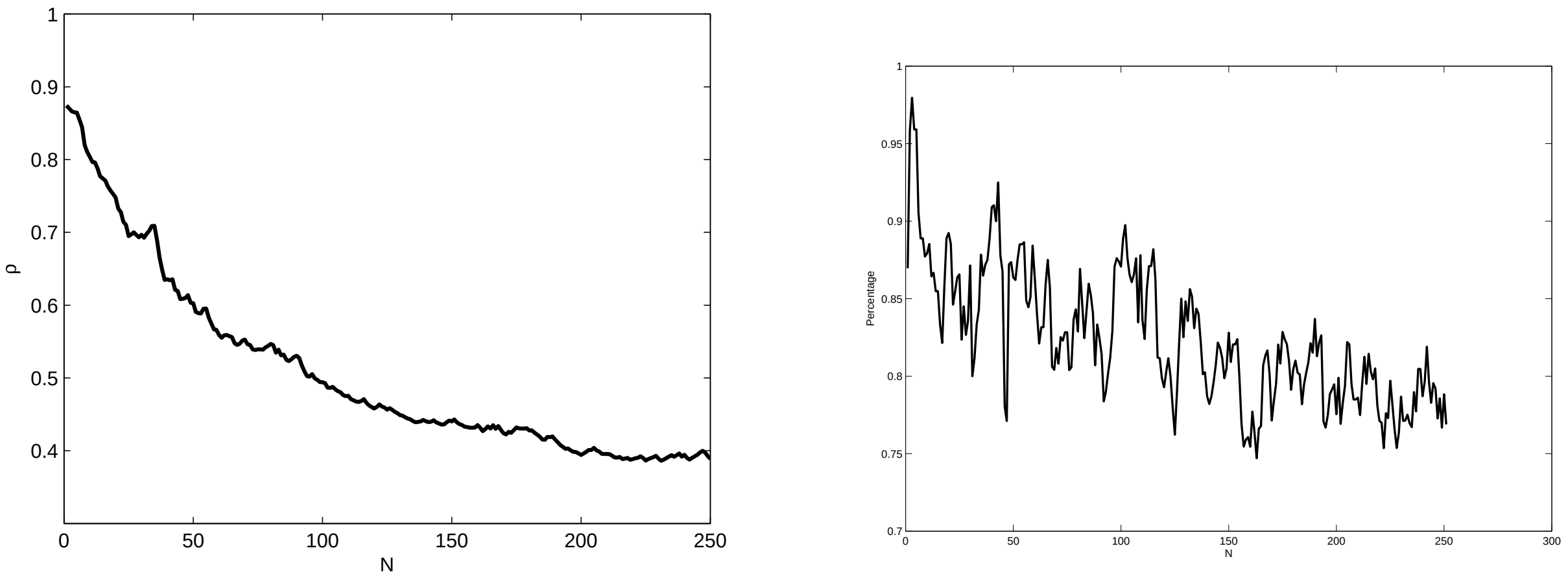

(a) Correlation coefficients for strong equivalence

(b) Percentage of points preserving local extremes for weak equivalence

Figure 8. (a), (b) shows the correlation coefficient and percentage of points preserving local extremes of the power spectrum change with growth of sequence length $N$ for Human gene AD169 sequences respectively using mapping which maps $\mathcal{A} = \{\texttt{A}, \texttt{T}, \texttt{G}, \texttt{C}\}$ to the standard basis of $\mathbb{R}^4$ and the RY rule

between the first mapping and any mapping obtained by rotating the second mapping.

## V. Conclusion

In this paper, we presented a novel framework for analysis of the equivalence of distinct numerical mappings of symbolic sequences undergoing a transformation by an operator. We introduced a strong equivalence property that demands perfect correlation between the transformations of distinct numerical representations. We also characterized the weak equivalence property which requires the preservation of the extrema in the transformation of the numerical representations. We studied the mapping equivalence theory for general operators by using Taylor's approximation. Moreover, we focused on first- and second-order operators such as the correlation function and Fourier transform. Furthermore, we derived the largest class of equivalent mappings which lead to consistent results when undergoing transformation by a class of operators. We demonstrated that rotation plays an important role in characterization of equivalence between distinct mappings. We subsequently derived a class of operators which is equivalent under rotations. We also introduced an abstract mapping model and extended the notion of equivalence to a more general algebraic structure. We presented simulations of the mathematical and statistical properties of genomic sequences in order to demonstrate the implications of the proposed mapping equivalence theory. Our results suggest that one of the reasons for inconsistency in the analysis of genomic data reported in the theoretical biology literature as well as many other related areas can be attributed to incompatibility of the numerical representation of symbolic data. For instance, we have shown that some of the mappings used for the representation of genomic data are incompatible and could have led to the contradictory conclusions reached in the analysis of long-range correlations of DNA sequences.

## ACKNOWLEDGMENT

The authors would like to thank Prof. Alan Willsky from the Massachusetts Institute of Technology for discussions that served as the impetus for the work presented in this paper.

## APPENDIX

*Proof of Theorem 3:*

Notice that $\mathbb{R}^n$ with inner product $< x, y >= x^T y$ is a Hilbert space. So for any linear bounded operator, $\exists y \in \mathbb{R}^n$, such that $\mathbf{\Phi}_l(x) = x^T y$. So $\mathbf{\Phi}_l(T(x)) = T^T x^T y$. $T$ is strongly equivalent, therefore $\mathbf{\Phi}_l(T(x)) = \lambda \mathbf{\Phi}_l(x) + c$ for some $\lambda$ and $c$. Then we have $T = \lambda I_{Nn\times Nn}$, i.e. $T$ is a trivial scaled identity transform. This finishes the proof of the first claim. We claim if a non-trivial operator $\mathbf{\Phi}_l(x)$ whose Taylor's expansion has no terms of order higher than or equal to three has a non-trivial linear *strongly equivalent* transformation, then it must only have the second order term and the constant term. We can always scale or add constant for the transformation to get a strongly equivalent result after transformation. So without loss of generality, we assume the result after the transformation is exactly the same as the previous one, i.e. If $\mathbf{\Phi}_l(x) = \frac{1}{2}\mathbf{x}^T \nabla^2 \mathbf{\Phi}_l(0)\mathbf{x}$ and $\mathbf{\Phi}_l(T(x)) = \mathbf{\Phi}_l(x)$, then we have

$$\frac{1}{2}\mathbf{x}^T \nabla^2 \mathbf{\Phi}_l(0)\mathbf{x} = \frac{1}{2}\mathbf{x}^T T^T (\nabla^2 \mathbf{\Phi}_l(0)) T\mathbf{x} \tag{34}$$

this equality holds for any $\mathbf{x} \in \mathbb{R}^{Nn\times 1}$. Therefore we have $T^T(\nabla^2 \mathbf{\Phi}_l(0))T = \nabla^2 \mathbf{\Phi}_l(0)$. $T$ is a rotation, i.e.

$$T = \begin{pmatrix} \mathbf{R}_{n\times n} & & & \\ & \mathbf{R}_{n\times n} & & \\ & & \ddots & \\ & & & \mathbf{R}_{n\times n} \end{pmatrix}_{Nn\times Nn} \tag{35}$$

Since $\mathbf{R}^T\mathbf{R} = I$, we have

$$(\nabla^2 \mathbf{\Phi}_l(0))T = T\nabla^2 \mathbf{\Phi}_l(0) \tag{36}$$

Because $\mathbf{R}^T\mathbf{R} = \mathbf{R}\mathbf{R}^T = I$, $T^T T = TT^T$, therefore $T$ is normal, $T$ is unitarily diagonalizable [37]. $\mathbf{R}$ is also normal. Therefore $\exists V$ unitary, such that $\mathbf{R} = V^H \Lambda' V$, where $\Lambda'$ is a diagonal matrix. Since $\mathbf{R}$ is real orthogonal, the eigenvalues of $\mathbf{R}$ are on the unit sphere $S^1$. Without loss of generality, we assume $\mathbf{R}$ has two eigenvalues, $1$ and $\mu$. Let the algebraic multiplicity of $1$ be $i$, then the algebraic multiplicity of $\mu$ is $n-i$. So we have

$$\Lambda' = \begin{pmatrix} I_{i\times i} & \\ & \mu I_{(n-i)\times(n-i)} \end{pmatrix} \tag{37}$$

Let $U = \begin{pmatrix} V & & \\ & \ddots & \\ & & V \end{pmatrix}$, $\Lambda = \begin{pmatrix} \Lambda' & & \\ & \ddots & \\ & & \Lambda' \end{pmatrix}$. From (36), we have $(\nabla^2 \mathbf{\Phi}_l(0))U^H \Lambda U = U^H \Lambda U \nabla^2 \mathbf{\Phi}_l(0)$. Therefore we have

$$U(\nabla^2 \mathbf{\Phi}_l(0))U^H \Lambda = \Lambda U \nabla^2 \mathbf{\Phi}_l(0) U^H \tag{38}$$

Let $\tilde{X} = U(\nabla^2 \mathbf{\Phi}_l(0))U^H$. We have $\tilde{X}\Lambda = \Lambda\tilde{X}$. By using the Jordan canonical form [38, Chapter VIII], we have that all $\tilde{X}$ which commutes with $\Lambda$ must have the form as follow:

$$\tilde{X} = \begin{pmatrix} A_{11} & 0 & A_{12} & 0 & \cdots & A_{1M} & 0 \\ 0 & B_{21} & 0 & B_{22} & \cdots & 0 & B_{2M} \\ A'_{12} & 0 & A_{31} & 0 & \cdots & A_{3M} & 0 \\ 0 & B'_{22} & 0 & B_{41} & \cdots & 0 & B_{4M} \\ \vdots & \vdots & \vdots & \vdots & \ddots & \ddots & \vdots \\ A'_{1M} & 0 & A'_{3M} & 0 & \vdots & A_{(2N-1)1} & 0 \\ 0 & B'_{2M} & 0 & B'_{4M} & \cdots & 0 & B_{(2N)1} \end{pmatrix} \tag{39}$$

Every non-zero submatrix in $\tilde{X}$ is an arbitrary upper triangular submatrix which has identical diagonal entries. All submatrices $A_{kl}$ and $A'_{kl}$ have size $i \times i$, all $B_{kl}$ and $B'_{kl}$ have size $(n-i)\times(n-i)$. Thus all $\nabla^2 \mathbf{\Phi}_l(0)$ satisfies (36) are of the form $\nabla^2 \mathbf{\Phi}_l(0) = U^H \tilde{X} U$. However, since the $\mathbf{\Phi}_l(x)$ is analytic, the Hessian must be symmetric. Therefore all the submatrices of $\tilde{X}$, $A_{kl}$ and $B_{kl}$ have the form $k_{kl}I$ and $A'_{kl}$ and $B'_{kl}$ have also the form $k_{kl}I$. Notice that this is for a given rotation. Since (36) holds for any given rotation, we have that

$$\nabla^2 \mathbf{\Phi}_l(0) \in Y = \bigcap_{U\, Unitary} \{U^H \tilde{X} U\} \tag{40}$$

The rotation is arbitrary, (37) should hold for any $i = 0, \ldots, n$. Claim that the $N \times N$ principal matrice $\begin{pmatrix} A_{(i)j} & \\ & B_{(i+1)j} \end{pmatrix}$ and $\begin{pmatrix} A'_{(i)j} & \\ & B'_{(i+1)j} \end{pmatrix}$ in $\tilde{X}$ must satisfy:

$$\begin{pmatrix} A_{(i)j} & \\ & B_{(i+1)j} \end{pmatrix} = k_{ij} I_{n \times n} \tag{41}$$

$$\begin{pmatrix} A'_{(i)j} & \\ & B'_{(i+1)j} \end{pmatrix} = k'_{ij} I_{n \times n} \tag{42}$$

where $i = 1, 3, 5, \ldots, (2N-1)$ and $j = 1, \ldots M$. Because we know that

$$\begin{pmatrix} A_{(i)j} & \\ & B_{(i+1)j} \end{pmatrix} = \begin{pmatrix} k_{ij} I_{i \times i} & \\ & k_{(i+1)j} I_{(n-i) \times (n-i)} \end{pmatrix} \tag{43}$$

but if $k_{ij} \neq k_{(i+1)j}$, then this implies $\{U^H \tilde{X} U\} \bigcap \{U^H \tilde{X}' U\} = \emptyset$, where (37) for $\tilde{X}'$ is of the form

$$\Lambda' = \begin{pmatrix} I_{(i+1) \times (i+1)} & \\ & \mu I_{(n-i-1) \times (n-i-1)} \end{pmatrix} \tag{44}$$

Since we know $Y$ is not empty, we get a contradiction here. Therefore (41) and (42) hold. If we choose $U = I$, we have $Y \subset \tilde{X}$, where for $\tilde{X}$, (41) and (42) hold. It's straightforward to check that such $\tilde{X}$ is commutable with $T$. Therefore $Y = \tilde{X}$. Finally we show that

$$\nabla^2 \mathbf{\Phi}_l(0) = \begin{pmatrix} k_{11} I_{n \times n} & k_{12} I_{n \times n} & \cdots & k_{1N} I_{n \times n} \\ k_{21} I_{n \times n} & k_{22} I_{n \times n} & \cdots & k_{2N} I_{n \times n} \\ \vdots & \vdots & \ddots & \vdots \\ k_{N1} I_{n \times n} & k_{N2} I_{n \times n} & \cdots & k_{NN} I_{n \times n} \end{pmatrix} \tag{45}$$

where $k_{ij} \in \mathbb{R}$ and $k_{ij} = k_{ji}$, $\forall i \neq j$. If we expand at any other point $x_0$, then

$$\frac{1}{2}(x-x_0)^T\nabla^2\mathbf{\Phi}_l(x_0)(x-x_0) = \frac{1}{2}x^T\nabla^2\mathbf{\Phi}_l(x_0)x-\frac{1}{2}x_0^T\nabla^2\mathbf{\Phi}_l(x_0)x-\frac{1}{2}x^T\nabla^2\mathbf{\Phi}_l(x_0)x_0+\frac{1}{2}x_0\nabla^2\mathbf{\Phi}_l(x_0)x_0 \tag{46}$$

The only second order term is $\frac{1}{2}x^T\nabla^2\mathbf{\Phi}_l(x_0)x$. Repeat the previous argument, we will have (10). ■